\documentclass[manuscript,screen]{acmart}

\AtBeginDocument{%
  }

\setcopyright{acmlicensed}
\copyrightyear{2024}
\acmYear{2024}
\acmDOI{XXXXXXX.XXXXXXX}

\acmISBN{978-1-4503-XXXX-X/18/06}

\usepackage{multirow}
\usepackage{booktabs}
\usepackage{caption,subcaption}
\usepackage{threeparttable}
\usepackage[noend,ruled,linesnumbered]{algorithm2e}
\usepackage{algorithmic}
\usepackage{color}
\usepackage{adjustbox}
\usepackage{tikz}
\usepackage{makecell}
\usepackage{tabularx}
\usepackage{bm}
\usepackage{amsmath}
\SetKwProg{Retract}{}{}{}
\usepackage{pifont}
\usepackage{xcolor}

\begin{document}

\title{Differentially Private Federated Learning: A Systematic Review}


\author{Jie Fu}
\authornote{This work is partially done under the supervision of Dr. Yang Cao at Hokkaido University.}
\email{jfu13@stevens.edu}
\affiliation{%
  \institution{Stevens Institute of Technology}
  \city{Hoboken}
  \country{USA}
}
\author{Yuan Hong}
\email{yuan.hong@uconn.edu}
\affiliation{%
  \institution{University of Connecticut}
  \city{Storrs}
  \country{USA}
}

\author{Xinpeng Ling}
\email{xpling@tongji.edu.cn}
\affiliation{%
  \institution{Tongji University}
  \city{Shanghai}
  \country{China}
}

\author{Leixia Wang}
\email{wangleixia@neu.edu.cn}
\affiliation{%
  \institution{Northeastern University}
  \city{Shenyang}
  \country{China}
}

\author{Xun Ran}
\email{qi-xun.ran@connect.polyu.hk}
\affiliation{%
  \institution{The Hong Kong Polytechnic University}
  \city{Hong Kong}
  \country{China}
}

\author{Zhiyu Sun}
\email{zysun@stu.ecnu.edu.cn}
\affiliation{%
  \institution{East China Normal University}
  \city{Shanghai}
  \country{China}
}

\author{Wendy Hui Wang}
\email{hui.wang@stevens.edu}
\affiliation{%
  \institution{Stevens Institute of Technology}
  \city{Hoboken}
  \country{USA}
}

\author{Zhili Chen}
\authornote{Corresponding author.}
\email{zhlchen@sei.ecnu.edu.cn}
\affiliation{%
  \institution{East China Normal University}
  \city{Shanghai}
  \country{China}
}

\author{Yang Cao}
\email{cao@c.titech.ac.jp}
\affiliation{%
  \institution{Institute of Science Tokyo}
  \city{Tokyo}
  \country{Japan}
}

\renewcommand{\shortauthors}{Jie Fu, et al.}

\begin{abstract}
In recent years, privacy and security concerns in machine learning have promoted trusted federated learning to the forefront of research. Differential privacy has emerged as the de facto standard for privacy protection in federated learning due to its rigorous mathematical foundation and provable guarantee. Despite extensive research on algorithms that incorporate differential privacy within federated learning, there remains an evident deficiency in systematic reviews that categorize and synthesize these studies.

Our work presents a systematic overview of the differentially private federated learning. Existing taxonomies have not adequately considered objects and level of privacy protection provided by various differential privacy models in federated learning. To rectify this gap, we propose a new taxonomy of differentially private federated learning based on definition and guarantee of various differential privacy models and federated scenarios. Our classification allows for a clear delineation of the protected objects across various differential privacy models and their respective neighborhood levels within federated learning environments. Furthermore, we explore the applications of differential privacy in federated learning scenarios. Our work provide valuable insights into privacy-preserving federated learning and suggest practical directions for future research.
\end{abstract}



\vspace{-0.3cm}
\keywords{differential privacy, federated learning, survey}
\vspace{-0.3cm}


\maketitle

\section{Introduction} \label{sec-introduction}
In the past decade, deep learning techniques have achieved remarkable success in many AI tasks~\cite{lundervold2019overview,chen2019gmail}.
In traditional deep learning frameworks, it is assumed that all pertinent training data is centralized under the governance of a singular, trustworthy entity. However, in real-world industrial contexts, data is often distributed across multiple independent parties, where a central trusted authority is often impractical. 
Additionally, legal restrictions such as CCPA~\cite{CCPA2018} or GDPR~\cite{cummings2018role}, along with business competition, may further limit the sharing of sensitive data. 

In response, federated learning (FL) has emerged as a promising collaborative learning infrastructure. 
In FL systems, participants retain their data locally, eliminating the need to share sensitive raw data. Instead, they contribute to a collective learning process by training models locally and sharing only the model parameters with a central server. The central server aggregates the updates and redistributes the refined model to all participants. However, even though the raw data are kept locally, adversaries may still infer sensitive data from the shared model parameters, posing severe privacy concerns. For example, the contents of raw data can be inverted from the model parameters~ \cite{zhu2019deep,nasr2019comprehensive,wang2019beyond}, or the membership information of the raw data can be inferred~\cite{song2017machine,melis2019exploiting}. To enhance the privacy of FL systems, several methods have been proposed based on homomorphic encryption ~\cite{damgaard2012multiparty} or secure multiparty computation (MPC)~\cite{mohassel2017secureml}. But these techniques require significant computational overheads and do not protect the final output of the computation, leaving them vulnerable to privacy breaches (e.g., inference attacks). One of the state-of-the-art (SOTA) paradigms to mitigate privacy risks in FL is differential privacy (DP)~ \cite{dwork2014algorithmic}. Many works \cite{feldman2020does, naseri2020local,jayaraman2019evaluating,stock2022defending} have demonstrated that by adding proper DP noise during the local training phase or uploaded model parameters, can prevent unintentional leakage of private raw data (e.g., via the membership inference attacks~\cite{song2017machine,melis2019exploiting}).

Currently, federated learning with differential privacy has attracted significant interests~\cite{naseri2020local,yang2023privatefl,xu2023learning,zhang2022understanding}. As the volume of publications on differentially private federated learning continues to grow, the task of summarizing and organizing this body of research has become both urgent and challenging. Despite the existence of surveys on differentially private federated learning~\cite{el2022differential,zhang2023systematic,farooqi2024differential,ren2024belt}, they do not provide a comprehensive, technical, and systematic analysis of differential privacy within the federated learning context. Table~\ref{tab:related-work} demonstrates the distinctions between our work and existing surveys. Specifically, our work differs from their in several key aspects: 1) We not only explore DP models in horizontal FL but also delve into vertical FL and transfer FL; 2) We propose a novel differentially private federated learning taxonomy based on the guarantees and definitions of DP, specifically categorizing DP\footnote{We note that previous literature often employs CDP (Centralized DP) in FL to distinguish it from LDP. The definition of CDP in those works is consistent with the definition of DP used in our work. In this work, we deliberately retain the original notation DP for differential privacy in the centralized setting, instead of introducing the additional symbol CDP.}, local DP (LDP), and the shuffle model in federated learning. Furthermore, we dissect the interconnections among these DP models and their respective privacy guarantees within the federated learning framework; 3) We elaborate on the relationship between DP, LDP, and the shuffle model in FL, and investigate different neighborhood levels for each DP models; 4) We provide a elaboration on different differential privacy composition mechanisms and their resulting effects; 5) We demonstrate the applications of differentially private federated learning in real-world scenarios.


\begin{table}[]
\footnotesize
\caption{The comparison between related works and ours.}
\label{tab:related-work}
\begin{tabular}{c|ccc|ccc|c|c|c}
\hline
\multirow{2}{*}{\textbf{\begin{tabular}[c]{@{}c@{}}Releated \\ works\end{tabular}}} & \multicolumn{3}{c|}{\textbf{Federated scenario}}                                                      & \multicolumn{3}{c|}{\textbf{DP model}}                                                         & \multirow{2}{*}{\textbf{\begin{tabular}[c]{@{}c@{}}Neighborhood \\ level\end{tabular}}} & \multirow{2}{*}{\textbf{\begin{tabular}[c]{@{}c@{}}Composition\\ mechanism\end{tabular}}} & \multirow{2}{*}{\textbf{Application}} \\ \cline{2-7}
                                                                                    & \multicolumn{1}{c|}{\textbf{Horizontal}} & \multicolumn{1}{c|}{\textbf{Vertical}} & \textbf{Transfer} & \multicolumn{1}{c|}{\textbf{CDP (DP)}} & \multicolumn{1}{c|}{\textbf{LDP}} & \textbf{Shuffle} &                                                                                         &                                                                                           &                                       \\ \hline
Farooqi et al.~\cite{farooqi2024differential}                                       & \multicolumn{1}{c|}{{\ding{51}}}         & \multicolumn{1}{c|}{}                  &                   & \multicolumn{1}{c|}{{\ding{51}}}  & \multicolumn{1}{c|}{}             &                        &                                                                                         &                                                                                           &                                       \\ \hline
Ren et al.~\cite{ren2024belt}                                                       & \multicolumn{1}{c|}{{\ding{51}}}         & \multicolumn{1}{c|}{}                  &                   & \multicolumn{1}{c|}{{\ding{51}}}  & \multicolumn{1}{c|}{{\ding{51}}}  &                        &                                                                                         &                                                                                           &                                       \\ \hline
El et al.~\cite{el2022differential}                                                 & \multicolumn{1}{c|}{{\ding{51}}}         & \multicolumn{1}{c|}{}                  &                   & \multicolumn{1}{c|}{{\ding{51}}}  & \multicolumn{1}{c|}{{\ding{51}}}  &                        &                                                                                         &                                                                                           &                                       \\ \hline
Zhang et al.~\cite{zhang2023systematic}                                             & \multicolumn{1}{c|}{{\ding{51}}}         & \multicolumn{1}{c|}{}                  &                   & \multicolumn{1}{c|}{{\ding{51}}}  & \multicolumn{1}{c|}{{\ding{51}}}  &                        &                                                                                         &                                                                                           & {\ding{51}}                           \\ \hline
Ours                                                                                & \multicolumn{1}{c|}{{\ding{51}}}         & \multicolumn{1}{c|}{{\ding{51}}}       & {\ding{51}}       & \multicolumn{1}{c|}{{\ding{51}}}  & \multicolumn{1}{c|}{{\ding{51}}}  & {\ding{51}}            & {\ding{51}}                                                                             & {\ding{51}}                                                                               & {\ding{51}}                           \\ \hline
\end{tabular}
\end{table}

\begin{figure*}[htb]
	\begin{center}
\includegraphics[width=0.7\linewidth]{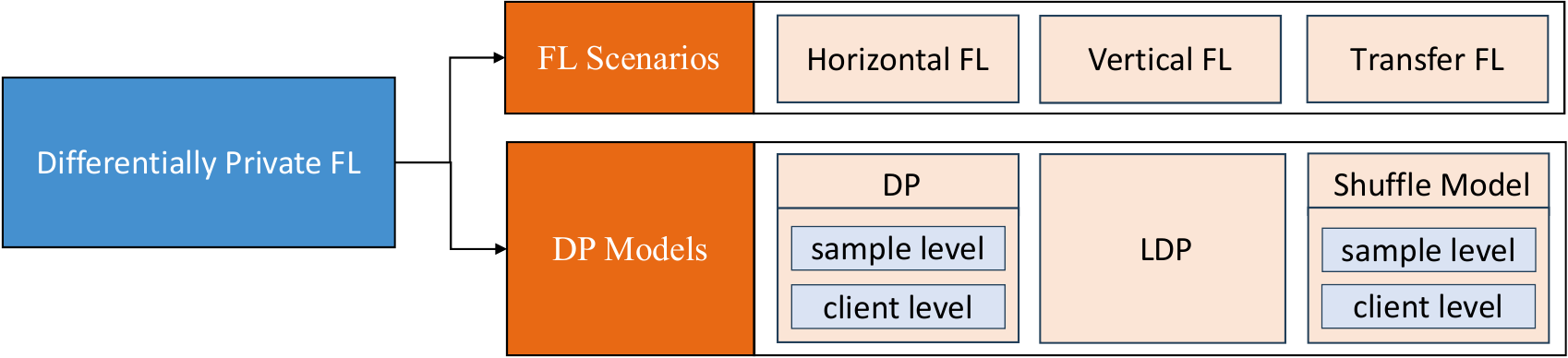}
		\caption{A new taxonomy of Differentially Private FL.}
		\label{fig: Taxonomy of DP-FL.}
	\end{center}
\vspace{-0.4cm}
\end{figure*}

As show in Figure~\ref{fig: Taxonomy of DP-FL.}, our survey classifies differentially private federated learning based on two aspects: FL scenarios and DP models. In FL scenarios, we classify the works to Horizontal FL, Vertical FL and Transfer FL by the data partitioning setting. We have encompassed all scenarios of federated learning. Although the majority of works on differential privacy for FL has been conducted in the context of Horizontal FL, as more privacy issues are exposed in vertical FL and transfer FL~\cite{pasquini2021unleashing,jagielski2024students}, research on DP models in them is becoming increasingly important. In DP models, We discuss differentially private federated learning from a new perspective, specifically, We distinguish between DP, LDP and shuffle model from the definition and privacy guarantee, which sets us apart from previous classifications based on the presence of a centralized trusted  server~\cite{zhang2023systematic,naseri2020local,yang2023privatefl,wei2021user}. We contend that their classification approach lacks precision. Federated learning is inherently a composite framework; for instance, in a cross-silo setting\footnote{According to the participating clients of FL, in cross-device setting, the number of clients is usually larger and consists of mobile devices. While clients are typically organizations or companies, and the number of clients is smaller in cross-silo setting~\cite{kairouz2021advances}.}, aside from a central server that aggregates model parameters, each client may also maintain a local server to manage data for training. Therefore, characterizing federated learning scenarios that lack a trusted central server for parameter aggregation solely as LDP is overly simplistic and inaccurate.
While these models can generally be grouped under the broad spectrum of DP, there are strict distinctions when considering the specific definitions and privacy guarantee. The differentiation and relation between them will be described in detail in Section~\ref{subsec:Relation between DP, LDP and shuffle}. Rigorous differentiation of DP models in differentially private federated learning, based on the definitions and privacy guarantee, is of significant importance. It not only clarifies the protection subjects of different DP models but also enhances the understanding of various implementation methods associated with these DP models. For DP and shuffle model, we further categorize into sample level and client level based on the definition of neighboring datasets -- corresponding to the protected objects \footnote{The definition of neighboring datasets does not exist in LDP.}.

In summary, our contributions are as follows:
\begin{itemize}
    \item We have reviewed the definitions and guarantee of DP, LDP and shuffle model, summarize the relaxation and differentiation of them. And we have further summarized the commonly used properties of DP models, privacy loss composition mechanisms, and perturbation methods in differentially private federated learning.

    \item We have broken away from the traditional classification method of differentially private FL, and explored the taxonomy from the definition and guarantee of differential privacy. We have provided a new, rigorous classification framework for differentially private FL.

    \item We have discussed and summarized over 70 recent articles on differentially private FL, as shown in Table~\ref{tab: An overview study of DP-FL}. We have clarified the protection targets of different DP models within federated learning. In addition to studying the use of differential privacy technology in horizontal FL, we have also showed how DP models to protect data privacy in vertical FL and transfer FL.
    
    \item We have summarized the applications of differentially private FL by data types and real-world implementation, and introduced related works in each domain.

    \item Based on the above research discussion, we have proposed 6 promising directions for future research.
\end{itemize}

The rest of the paper is organized as follows. As shown in Table \ref{tab: An overview study of DP-FL}, we will first introduce the research status of three DP models in horizontal federated learning, followed by the three DP models in vertical federated learning, and finally the three DP models in transfer federated learning. Specially, Section~\ref{sec-preliminary} presents three FL settings and three DP models. In Section~\ref{sec-Horizontal}, we discuss the various DP models in horizontal FL. Section~\ref{sec-VerticalandTransfer} shows the differential privacy techniques in vertical FL and transfer FL. Section~\ref{sec-application} further explores the applications of differentially private federated learning in the real world. Open challenges and future directions are introduced in Section~\ref{sec-future direction}, followed the conclusion in Section~\ref{sec-conclusion}.

\begin{table*}[]
\caption{An overview study of Differentially Private FL.}
\label{tab: An overview study of DP-FL}
\tiny
\begin{tabular}{|c|c|c|c|c|c|c|c|c|c|c|c|}
\hline
\textbf{\begin{tabular}[c]{@{}c@{}}Federated\\  Scenario\end{tabular}} & \textbf{Publications}                                & \textbf{Year} & \textbf{\begin{tabular}[c]{@{}c@{}}DP \\ Model\end{tabular}} & \textbf{\begin{tabular}[c]{@{}c@{}}Neighborhood\\ Level\end{tabular}} & \textbf{\begin{tabular}[c]{@{}c@{}}Perturbation \\ Mechanism\end{tabular}} & $\textbf{CM}^1$ & \textbf{\begin{tabular}[c]{@{}c@{}}Downsteam\\  Tasks\end{tabular}} & \textbf{$\begin{tabular}[c]{@{}c@{}}Model \\ Archiecture\end{tabular}^2$} & \textbf{\begin{tabular}[c]{@{}c@{}}Clients\\  Number\end{tabular}} & \textbf{$\epsilon$} & \textbf{$\delta$}                  \\ \hline
                                                                       & Chen et al.\cite{chen2024differentially}             & 2024          &                                                              &                                                                       & Gaussian                                                                   & tCDP                                                                     & Classification                                                      & LR, Shallow CNN                                                       & 100                                                                & 0.3                 & $10^{-2}$                          \\ \cline{2-3} \cline{6-12} 
                                                                       & malekmohammadi et al. \cite{malekmohammadi2024noise} & 2024          &                                                              &                                                                       & Gaussian                                                                   & AC                                                                       & Classification                                                      & CNN                                                                   & [20,60]                                                            & [0.5,5]             & $10^{-4}$                          \\ \cline{2-3} \cline{6-12} 
                                                                       & Liu et al.\cite{liu2024cross}                        & 2024          &                                                              &                                                                       & Gaussian                                                                   & RDP                                                                      & Classification                                                      & CNN                                                                   & 10                                                                 & [0.1,10]            & $10^{-3}$                          \\ \cline{2-3} \cline{6-12} 
                                                                       & Ling et al. \cite{ling2024ali}                       & 2024          &                                                              &                                                                       & Gaussian                                                                   & RDP                                                                      & Classification                                                      & Shallow CNN                                                           & 10                                                                 & [1.5,5.5]           & $10^{-5}$                          \\ \cline{2-3} \cline{6-12} 
                                                                       & Xiang et al. \cite{xiang2023practical}               & 2023          &                                                              &                                                                       & Gaussian                                                                   & MA                                                                       & Classification                                                      & Shallow CNN,LSTM                                                      & [10,20]                                                            & [0.12,2]            & $[10^{-2},10^{-5}]$                \\ \cline{2-3} \cline{6-12} 
                                                                       & Ruan et al. \cite{ruan2023private}                   & 2023          &                                                              &                                                                       & Gaussian                                                                   & RDP                                                                      & Classification                                                      & Shallow CNN, LSTM                                                     & [3,10]                                                             & [0.25,2]            & $[10^{-4},10^{-5}]$                \\ \cline{2-3} \cline{6-12} 
                                                                       & Noble et al. \cite{noble2022differentially}          & 2022          &                                                              &                                                                       & Gaussian                                                                   & RDP                                                                      & Classification                                                      & Shallow CNN                                                           & 10                                                                 & [3,13]              & $10^{-6}$                          \\ \cline{2-3} \cline{6-12} 
                                                                       & Fu et al. \cite{fu2022adap}                          & 2022          &                                                              &                                                                       & Gaussian                                                                   & RDP                                                                      & Classification                                                      & Shallow CNN                                                           & 10                                                                 & [2,6]               & $10^{-5}$                          \\ \cline{2-3} \cline{6-12} 
                                                                       & Li et al. \cite{li2022soteriafl}                     & 2022          &                                                              &                                                                       & Gaussian                                                                   & MA                                                                       & Classification                                                      & LR, Shallow CNN                                                       & 10                                                                 & [1,16]              & $10^{-3}$                          \\ \cline{2-3} \cline{6-12} 
                                                                       & Ryu et al. \cite{ryu2022differentially}              & 2022          &                                                              &                                                                       & Gaussian                                                                   & AC                                                                       & Classification                                                      & LR                                                                    & [10,195]                                                           & [0.05,5]            & $10^{-6}$                          \\ \cline{2-3} \cline{6-12} 
                                                                       & Wei et al. \cite{wei2021user}                        & 2021          &                                                              &                                                                       & Gaussian                                                                   & MA                                                                       & Classification                                                      & Shallow CNN                                                           & 50                                                                 & [4,20]              & $10^{-3}$                          \\ \cline{2-3} \cline{6-12} 
                                                                       & Liu et al. \cite{liu2021projected}                   & 2021          &                                                              &                                                                       & Gaussian                                                                   & GDP                                                                      & Classification                                                      & Shallow CNN                                                           & 100                                                                & [10,100]            & $10^{-3}$                          \\ \cline{2-3} \cline{6-12} 
                                                                       & Zheng et al. \cite{zheng2021federated}               & 2021          &                                                              &                                                                       & Gaussian                                                                   & GDP                                                                      & Classification                                                      & Shallow CNN                                                           & 100                                                                & [10,100]            & $10^{-3}$                          \\ \cline{2-3} \cline{6-12} 
                                                                       & Huang et al. \cite{huang2020dp}                      & 2020          &                                                              &                                                                       & Gaussian, Laplace                                                          & AC                                                                       & Classification                                                      & Shallow CNN                                                           & {10,100,1000}                                                      & [0.2,8]             & $[10^{-2},10^{-5}]$                \\ \cline{2-3} \cline{6-12} 
                                                                       & Wei et al. \cite{wei2020federated}                   & 2020          &                                                              &                                                                       & Gaussian                                                                   & MA                                                                       & Classification                                                      & Shallow CNN,LSTM                                                      & [10,20]                                                            & [0.12,2]            & $[10^{-2},10^{-5}]$                \\ \cline{2-3} \cline{6-12} 
                                                                       & Huang et al. \cite{huang2019dp}                      & 2019          &                                                              & \multirow{-16}{*}{SL}                                                 & Gaussian                                                                   & AC                                                                       & Regression                                                          & LR                                                                    & -                                                                  & [0.01,0.2]          & $[10^{-3},10^{-6}]$                \\ \cline{2-3} \cline{5-12} 
                                                                       & Yang et al. \cite{yang2023dynamic}                   & 2023          &                                                              &                                                                       & Gaussian                                                                   & RDP                                                                      & Classification                                                      & Shallow CNN                                                           & 50                                                                 & [2,16]              & $10^{-3}$                          \\ \cline{2-3} \cline{6-12} 
                                                                       & Xu et al. \cite{xu2023learning}                      & 2023          &                                                              &                                                                       & Gaussian                                                                   & RDP                                                                      & Classification                                                      & ResNet-50                                                             & [1262,9896000]                                                     & [10,20]             & $10^{-7}$                          \\ \cline{2-3} \cline{6-12} 
                                                                       & Shi et al. \cite{shi2023make}                        & 2023          &                                                              &                                                                       & Gaussian                                                                   & RDP                                                                      & Classification                                                      & ResNet-18                                                             & 500                                                                & [4,10]              & $\frac{1}{500}$                    \\ \cline{2-3} \cline{6-12} 
                                                                       & Zhang et al. \cite{zhang2022understanding}           & 2022          &                                                              &                                                                       & Gaussian                                                                   & MA                                                                       & Classification                                                      & Shallow CNN, ResNet-18                                                & 1920                                                               & [1.5,5]             & $10^{-5}$                          \\ \cline{2-3} \cline{6-12} 
                                                                       & Cheng et al. \cite{cheng2022differentially}          & 2022          &                                                              &                                                                       & Gaussian                                                                   & MA                                                                       & Classification                                                      & Shallow CNN, ResNet-18                                                & 3400                                                               & [2,8]               & $\frac{1}{3400}$                   \\ \cline{2-3} \cline{6-12} 
                                                                       & Bietti et al. \cite{bietti2022personalization}       & 2022          &                                                              &                                                                       & Gaussian                                                                   & MA                                                                       & Classification                                                      & Shallow CNN                                                           & 1000                                                               & [0.1,1000]          & $10^{-4}$                          \\ \cline{2-3} \cline{6-12} 
                                                                       & Andrew et al. \cite{andrew2021differentially}        & 2021          &                                                              &                                                                       & Gaussian                                                                   & RDP                                                                      & Classification                                                      & Shallow CNN                                                           & [500,342000]                                                       & [0.035,5]           & $[\frac{1}{500},\frac{1}{342000}]$ \\ \cline{2-3} \cline{6-12} 
                                                                       & Mamahan et al. \cite{mcmahan2017learning}            & 2017          &                                                              &                                                                       & Gaussian                                                                   & MA                                                                       & Classification                                                      & LSTM                                                                  & [100,763430]                                                       & [2.0,4.6]           & $10^{-9}$                          \\ \cline{2-3} \cline{6-12} 
                                                                       & Geyer et al. \cite{geyer2017differentially}          & 2017          &                                                              & \multirow{-9}{*}{CL}                                                  & Gaussian                                                                   & MA                                                                       & Classification                                                      & Shallow CNN                                                           & {100, 1000, 10000}                                                 & 8                   & $[10^{-3},10^{-6}]$                \\ \cline{2-3} \cline{5-12} 
                                                                       & Chen et al. \cite{chen2022fundamental}               & 2022          &                                                              &                                                                       & Discrete Gaussian                                                          & RDP                                                                      & Classification                                                      & Shallow CNN                                                           & [100,1000]                                                         & [0,10]              & $10^{-2}$                          \\ \cline{2-3} \cline{6-12} 
                                                                       & Chen et al. \cite{chen2022poisson}                   & 2022          &                                                              &                                                                       & Poisson Binomial                                                           & RDP                                                                      & Classification                                                      & LR                                                                    & 1000                                                               & [0.5,6]             & $10^{-5}$                          \\ \cline{2-3} \cline{6-12} 
                                                                       & Wang et al. \cite{wang2020d2p}                       & 2020          &                                                              &                                                                       & Discrete Gaussian                                                          & RDP                                                                      & Classification                                                      & Shallow CNN                                                           & 100K                                                               & [2,4]               & $10^{-5}$                          \\ \cline{2-3} \cline{6-12} 
                                                                       & Stevens et al. \cite{stevens2022efficient}           & 2022          &                                                              &                                                                       & LWE                                                                        & RDP                                                                      & Classification                                                      & Shallow CNN                                                           & [500,1000]                                                         & [2,8]               & $10^{-5}$                          \\ \cline{2-3} \cline{6-12} 
                                                                       & Kairouz et al. \cite{kairouz2021distributed}         & 2021          &                                                              &                                                                       & Discrete Gaussian                                                          & zCDP                                                                     & Classification                                                      & Shallow CNN                                                           & 3400                                                               & [3,10]              & $\frac{1}{3400}$                   \\ \cline{2-3} \cline{6-12} 
                                                                       & Agarwal et al. \cite{agarwal2021skellam}             & 2021          &                                                              &                                                                       & Skellam                                                                    & RDP                                                                      & Classification                                                      & Shallow CNN                                                           & 1000k                                                              & [5,20]              & $10^{-6}$                          \\ \cline{2-3} \cline{6-12} 
                                                                       & Kerkouche et al. \cite{kerkouche2021compression}     & 2021          &                                                              &                                                                       & Gaussian                                                                   & MA                                                                       & Classification                                                      & Shallow CNN                                                           & [5011,6000]                                                        & [0.5,1]             & $10^{-5}$                          \\ \cline{2-3} \cline{6-12} 
                                                                       & Agarwal et al. \cite{agarwal2018cpsgd}               & 2018          &                                                              & \multirow{-8}{*}{CL with SA}                                          & Binomial                                                                   & AC                                                                       & Classification                                                      & LR                                                                    & 25M                                                                & [2,4]               & $10^{-9}$                          \\ \cline{2-3} \cline{5-12} 
                                                                       & Naseri et al. \cite{naseri2020local}                 & 2020          &                                                              & SL, CL                                                                & Gaussian                                                                   & RDP                                                                      & Classification                                                      & Shallow CNN,LSTM                                                      & [100,660120]                                                       & [1.2,10.7]          & $10^{-5}$                          \\ \cline{2-3} \cline{5-12} 
                                                                       & Yang et al. \cite{yang2023privatefl}                 & 2023          & \multirow{-35}{*}{DP}                                        & SL, CL, CL with SA                                                    & Gaussian, Skellam                                                          & RDP                                                                      & Classification                                                      & Shallow CNN                                                           & [40,500]                                                           & [2,8]               & $10^{-3}$                          \\ \cline{2-12} 
                                                                       & Triastcyn et al. \cite{triastcyn2019federated}       & 2019          & Bayesian DP                                                  & SL, CL                                                                & Gaussian                                                                   & RDP                                                                      & Classification                                                      & ResNet-50                                                             & [100,10000]                                                        & [0.2,4]             & $[10^{-3},10^{-6}]$                \\ \cline{2-12} 
                                                                       & Zhang et al. \cite{zhang2024dynamic}                 & 2024          &                                                              &                                                                       & Gaussian                                                                   & zCDP                                                                     & Classification                                                      & LR                                                                    & 20                                                                 & 1                   & $10^{-4}$                          \\ \cline{2-3} \cline{6-12} 
                                                                       & Varun et al. \cite{Varun24SSRFL}                     & 2024          &                                                              &                                                                       & SRR                                                                        & BC                                                                       & Classification                                                      & Shallow CNN                                                           & 100                                                                & [1,10]              & 0                                  \\ \cline{2-3} \cline{6-12} 
                                                                       & Zhang et al. \cite{zhang2023pfldp}                   & 2023          &                                                              &                                                                       & Gaussian                                                                   & AC                                                                       & Classification                                                      & LR, Shallow CNN                                                       & 100                                                                & [3,30]              & -                                  \\ \cline{2-3} \cline{6-12} 
                                                                       & Wang et al. \cite{wang2023ppefl}                     & 2023          &                                                              &                                                                       & EM, DMP-UE                                                                 & BC                                                                       & Classification                                                      & Shallow CNN                                                           & [10,50]                                                            & [0.1,1]             & 0                                  \\ \cline{2-3} \cline{6-12} 
                                                                       & Jiang et al. \cite{jiang2022signds}                  & 2023          &                                                              &                                                                       & EM                                                                         & BC                                                                       & Classification                                                      & Shallow CNN                                                           & [100,750]                                                          & [0.5,12]            & 0                                  \\ \cline{2-3} \cline{6-12} 
                                                                       & Li et al. \cite{li2022fedta}                         & 2022          &                                                              &                                                                       & Laplace                                                                    & BC                                                                       & Classification                                                      & Shallow CNN                                                           & 100                                                                & 78.5                & 0                                  \\ \cline{2-3} \cline{6-12} 
                                                                       & Lian et al. \cite{lian2022webfed}                    & 2022          &                                                              &                                                                       & Laplace                                                                    & BC                                                                       & Classification                                                      & Shallow CNN                                                           & 5                                                                  & [3,6]               & 0                                  \\ \cline{2-3} \cline{6-12} 
                                                                       & Mahawaga et al. \cite{mahawaga2022local}             & 2022          &                                                              &                                                                       & RAPPOR                                                                     & BC                                                                       & Classification                                                      & Shallow CNN                                                           & [2,100]                                                            & [0.5,10]            & 0                                  \\ \cline{2-3} \cline{6-12} 
                                                                       & Wang et al. \cite{wang2022safeguarding}              & 2022          &                                                              &                                                                       & RAPPOR                                                                     & BC                                                                       & Classification                                                      & LR                                                                    & [500,1800]                                                         & [0.1,10]            & 0                                  \\ \cline{2-3} \cline{6-12} 
                                                                       & Zhao et al. \cite{zhao2022privacy}                   & 2022          &                                                              &                                                                       & Adaptive-Harmony                                                           & BC                                                                       & Classification                                                      & Shallow CNN                                                           & 200                                                                & [1,10]              & 0                                  \\ \cline{2-3} \cline{6-12} 
                                                                       & Sun et al. \cite{sun2021ldp}                         & 2021          &                                                              &                                                                       & Adaptive-Duchi                                                             & BC                                                                       & Classification                                                      & Shallow CNN                                                           & [100,500]                                                          & [1,5]               & 0                                  \\ \cline{2-3} \cline{6-12} 
                                                                       & Yang et al. \cite{yang2021federated}                 & 2021          &                                                              &                                                                       & Laplace                                                                    & BC                                                                       & Classification                                                      & Shallow CNN                                                           & [200,1000]                                                         & [1,10]              & 0                                  \\ \cline{2-3} \cline{6-12} 
                                                                       & Wang et al. \cite{wang2020federated}                 & 2020          &                                                              &                                                                       & RRP                                                                        & AC                                                                       & Topic Modeling                                                      & LDA                                                                   & 150                                                                & [5,8]               & [0.05,0.5]                         \\ \cline{2-3} \cline{6-12} 
                                                                       & Zhao et al. \cite{zhao2020local}                     & 2020          &                                                              &                                                                       & Three output, PM-SUB                                                       & BC                                                                       & Classification                                                      & LR, SVM                                                               & 4M                                                                 & [0.5,4]             & 0                                  \\ \cline{2-3} \cline{6-12} 
                                                                       & Liu et al. \cite{liu2020fedsel}                      & 2020          &                                                              &                                                                       & RR, PM                                                                     & BC                                                                       & Classification                                                      & LR, SVM                                                               & 4W-10W                                                             & [0.5,16]            & 0                                  \\ \cline{2-3} \cline{6-12} 
                                                                       & Wang et al. \cite{wang2019collecting}                & 2019          & \multirow{-16}{*}{LDP}                                       & \multirow{-16}{*}{-}                                                  & PM                                                                         & BC                                                                       & Classification                                                      & LR, SVM                                                               & 4M                                                                 & [0.5,4]             & 0                                  \\ \cline{2-12} 
                                                                       & Truex et al. \cite{truex2020ldp}                     & 2020          & Condensed LDP                                                & -                                                                     & EM                                                                         & BC                                                                       & Classification                                                      & Shallow CNN                                                           & 50                                                                 & 1                   & 0                                  \\ \cline{2-12} 
                                                                       & Liu et al. \cite{liu2023echo}                        & 2023          &                                                              &                                                                       & Clipped-Laplace,Shuffle                                                    & AC                                                                       & Classification                                                      & LR                                                                    & 10000                                                              & 25.6                & $10^{-8}$                          \\ \cline{2-3} \cline{6-12} 
                                                                       & Liew et al. \cite{liew2022shuffled}                  & 2022          &                                                              &                                                                       & Harmony,Shuffle                                                            & RDP                                                                      & Classification                                                      & Shallow CNN                                                           & [50000,60000]                                                      & [2.8]               & -                                  \\ \cline{2-3} \cline{6-12} 
                                                                       & Liu et al. \cite{liu2021flame}                       & 2021          &                                                              & \multirow{-3}{*}{CL}                                                  & Laplace,Shuffle                                                            & BC, AC                                                                   & Classification                                                      & LR                                                                    & 1000                                                               & 4.696               & $5 * 10^{-6}$                      \\ \cline{2-3} \cline{5-12} 
                                                                       & Chen et al. \cite{chen2023generalized}               & 2023          &                                                              &                                                                       & Duchi,Shuffle                                                              & GDP                                                                      & Classification                                                      & Shallow CNN                                                           & 100                                                                & [0.5, 100]          & $10^{-5}$                          \\ \cline{2-3} \cline{6-12} 
\multirow{-58}{*}{Horizontal}                                          & Girgis et al. \cite{girgis2021shuffled}              & 2021          & \multirow{-5}{*}{Shuffle DP}                                 & \multirow{-2}{*}{SL}                                                  & Laplace,Shuffle                                                            & AC                                                                       & Classification                                                      & Shallow CNN                                                           & 60000                                                              & [1,10]              & $10^{-5}$                          \\ \hline
                                                                       & Takahashi et al. \cite{takahashi2023eliminating}     & 2023          &                                                              &                                                                       & KRR                                                                        & BC                                                                       & Classification                                                      & GBDT                                                                  & 3                                                                  & [0.1,2.0]           & -                                  \\ \cline{2-3} \cline{6-12} 
                                                                       & Yang et al. \cite{yang2022differentially}            & 2022          & \multirow{-2}{*}{Label DP}                                   &                                                                       & Laplace, KRR                                                               & BC                                                                       & Classification                                                      & Shallow CNN                                                           & 2                                                                  & 1                   & 0                                  \\ \cline{2-4} \cline{6-12} 
                                                                       & Oh et al. \cite{oh2022differentially}                & 2022          &                                                              & \multirow{-3}{*}{SL}                                                  & Gaussian                                                                   & RDP                                                                      & Classification                                                      & VGG-16                                                                & 10                                                                 & [1,40]              & -                                  \\ \cline{2-3} \cline{5-12} 
                                                                       & Chen et al. \cite{chen2020vafl}                      & 2020          &                                                              &                                                                       & Gaussian                                                                   & GDP                                                                      & Classification                                                      & Shallow CNN                                                           & [3,8]                                                              & -                   & -                                  \\ \cline{2-3} \cline{6-12} 
                                                                       & Wang et al. \cite{wang2020hybrid}                    & 2020          &                                                              &                                                                       & Gaussian                                                                   & AC                                                                       & Classification                                                      & Shallow CNN                                                           & 2                                                                  & [0.001, 10]         & $10^{-2}$                          \\ \cline{2-3} \cline{6-12} 
                                                                       & Wu et al. \cite{wu2020privacy}                       & 2020          & \multirow{-4}{*}{DP}                                         & \multirow{-3}{*}{CL}                                                  & Laplace                                                                    & BC                                                                       & Classification                                                      & GBDT                                                                  & [2,10]                                                             & -                   & -                                  \\ \cline{2-12} 
                                                                       & Mao et al. \cite{mao2022secure}                      & 2022          &                                                              &                                                                       & Laplace, RR                                                                & BC                                                                       & Classification                                                      & Shallow CNN                                                           & 5                                                                  & [0.1,4.0]           & 0                                  \\ \cline{2-3} \cline{6-12} 
                                                                       & Tian et al. \cite{tian2020federboost}                & 2020          & \multirow{-2}{*}{LDP}                                        & \multirow{-2}{*}{-}                                                   & RR                                                                         & BC                                                                       & Classification                                                      & GBDT                                                                  & 3                                                                  & 4                   & 0                                  \\ \cline{2-12} 
\multirow{-9}{*}{Vertical}                                             & Li et al. \cite{li2022opboost}                       & 2022          & Condensed LDP                                                & -                                                                     & Discrete Laplace                                                           & BC                                                                       & Classification                                                      & GBDT                                                                  & 2                                                                  & [0.64,2.56]         & 0                                  \\ \hline
                                                                       & Wan et al. \cite{wan2023fedpdd}                      & 2023          &                                                              &                                                                       & Gaussian                                                                   & AC                                                                       & Recommendection                                                     & DeepFM                                                                & 2                                                                  & [0.05, 10]          & -                                  \\ \cline{2-3} \cline{6-12} 
                                                                       & Hoech et al. \cite{hoech2022fedauxfdp}               & 2022          &                                                              &                                                                       & Gaussian                                                                   & AC                                                                       & Classification                                                      & Resnet-18                                                             & 20                                                                 & [0.1,0.5]           & -                                  \\ \cline{2-3} \cline{6-12} 
                                                                       & Tian et al. \cite{tian2022seqpate}                   & 2022          &                                                              &                                                                       & Gaussian                                                                   & GDP                                                                      & Text Generation                                                     & GPT-2                                                                 & 2000                                                               & [3,5]               & $10^{-6}$                          \\ \cline{2-3} \cline{6-12} 
                                                                       & Sun et al. \cite{sun2020federated}                   & 2020          &                                                              &                                                                       & Random Sampling                                                            & AC                                                                       & Classification                                                      & Shallow CNN                                                           & 6                                                                  & [0.003,0.65]        & [0.006,0.65]                       \\ \cline{2-3} \cline{6-12} 
                                                                       & Papernot er al. \cite{papernot2018scalable}          & 2018          &                                                              &                                                                       & Gaussian                                                                   & RDP                                                                      & Classification                                                      & Resnet-18                                                             & 2                                                                  & [0.59,8.03]         & $10^{-8}$                          \\ \cline{2-3} \cline{6-12} 
                                                                       & Papernot er al. \cite{papernot2017semi}              & 2017          &                                                              & \multirow{-6}{*}{SL}                                                  & Laplace                                                                    & MA                                                                       & Classification                                                      & Shallow CNN                                                           & 2                                                                  & [2.04,8.19]         & $[10^{-5},10^{-6}]$                \\ \cline{2-3} \cline{5-12} 
                                                                       & Dodwadmath et al. \cite{dodwadmath2022preserving}    & 2022          &                                                              &                                                                       & Laplace                                                                    & MA                                                                       & Classification                                                      & Shallow CNN                                                           & 10                                                                 & [11.75,20]          & $10^{-5}$                          \\ \cline{2-3} \cline{6-12} 
                                                                       & Pan et al. \cite{pan2021fl}                          & 2021          & \multirow{-8}{*}{DP}                                         & \multirow{-2}{*}{CL}                                                  & Gaussian                                                                   & RDP                                                                      & Classification                                                      & Resnet-18                                                             & 100                                                                & [0.95,9.03]         & -                                  \\ \cline{2-12} 
\multirow{-9}{*}{Transfer}                                            & Qi et al. \cite{qi2023differentially}                & 2023          & LDP                                                          & -                                                                     & KRR                                                                        & BC                                                                       & Classification                                                      & Shallow CNN                                                           & [2,5]                                                              & [2,7]               & 0                                  \\ \hline
\end{tabular}
	\begin{tablenotes}
		\footnotesize
		\item[1] 1. CM=Composition Mechanism, BC=Basic Sequential Composition Theory, AC=Advanced Sequential Composition Theory. 
  		\item[2] 2. LR=Logistic Regression, SVM=Support Vector Machine, GBDT=Gradient Boosting Decision Tree. 
    \end{tablenotes}
\end{table*}
\vspace{-0.2cm}
\section{FL and DP Models} \label{sec-preliminary}
In this section, we will describe three FL scenarios, and explain the definition and relation of three DP models. Additionally, we provide a discussion on popular loss composition mechanisms in DP models and fundamental used perturbation mechanisms for differentially private FL in Section~\ref{sec:Loss Composition Mechanisms in DP} and Section~\ref{sec:Fundamental Mechanisms for differentially private FL}, respectively.

\subsection{Federated Learning}

Federated learning (FL) is a distributed machine learning paradigm where multiple clients collaboratively train a global model under the coordination of a central server, without sharing their raw data. Based on how data is distributed among participants~\cite{li2021survey}, FL can be categorized as follows.

\subsubsection{Horizontal Federated Learning (HFL)}
HFL applies to scenarios where participants hold different data samples but share the same feature space. Its objective is to collaboratively train a global model $\mathbf{w}^\star$ by aggregating model updates from all clients to minimize a weighted global loss function $F(\mathbf{w})$:
\begin{align}
\mathbf{w}^\star \triangleq \min_{\mathbf{w}} F(\mathbf{w}), \,
\text{where} \, F(\mathbf{w}_t) \triangleq \sum_{k=1}^K p_k L_k(\mathbf{w}_t^k).
\end{align}
Here, $p_k$ is the aggregate weight of client $k$, $L_k(\cdot)$ is its loss function, and $\mathbf{w}_t^k$ is the model uploaded by client $k$ at communication round $t$.

\subsubsection{Vertical Federated Learning (VFL)}
VFL applies to scenarios where participants share the same data sample IDs but have different feature spaces. The goal is to jointly train a global model $\mathbf{w}^\star$ by combining features from all parties~\cite{yang2019federated}. Typically, one party (the server) holds the data labels, while the others (the clients) only have feature information.
\begin{align}
\mathbf{w}^{\star} &\triangleq \min_{\mathbf{w}} F\left(\mathbf{w} \right), \, \text{where } \, F\left(\mathbf{w}\right) \triangleq\frac{1}{N} \sum_{i=1}^{N} L_K(\mathbf{w}_K ; h_1\left(x_{i, 1}, \mathbf{w}_1\right), \ldots,
h_K\left(x_{i, K}, \mathbf{w}_K\right), y_{i, K}).
\end{align}
Here, $L_K(\cdot)$ is the loss function of the server with the label information.

\subsubsection{Transfer Federated Learning (TFL)}
TFL refers to FL scenarios with limited or no overlap in feature or sample spaces among participants~\cite{liu2020secure}. In this setting, clients with source domain datasets and a server with a target domain dataset jointly train a model to generalize to the target domain while preserving data privacy. If the target domain lacks labels, it often requires using source domain data to generate pseudo-labels for the target domain.

\subsection{DP Models} \label{subsec-dp}
Currently, there are three main DP models, which are DP, LDP, and shuffle model. They have different definitions and privacy guarantees, but there is a connection between them.
The introduction and relation to these three DP models and some properties of DP will be presented as follows. 

\subsubsection{DP} 
Differential privacy (DP)~\cite{dwork2006calibrating} is a formal definition of data privacy, primarily used in centralized settings with a trusted server. It protects against privacy attacks by adding noise to the output of statistical queries. This mechanism ensures that the inclusion or exclusion of any single data record does not cause a statistically significant change in the output, thus safeguarding individual privacy.

\vspace{-0.2cm}
\begin{definition}\label{def-Differential Privacy}
    ({\bf DP~\cite{dwork2014algorithmic}}). The randomized algorithm $\mathcal{A}:\mathbb{X}^n\to\mathbb{Y}$ satisfies $(\epsilon,\delta)$-DP if any two neighboring datasets $D$ and $D'$ that differ in only a single entry and $D\simeq D^{\prime} \in \mathbb{X}^n$, we have 
    \begin{align*}
        \forall S\subseteq\mathbb{Y}:\mathrm{Pr}[\mathcal{A}(D)\in S]\leq e^{\epsilon}\mathrm{Pr}[\mathcal{A}(D^{\prime})\in S]+\delta.
    \end{align*}
\end{definition}
\vspace{-0.2cm}

Here, $\epsilon > 0$ controls the level of privacy guarantee in the worst case. The smaller $\epsilon$, the stronger the privacy level is. The factor $\delta \ge 0$ is the failure probability that the property does not hold. In practice, the value of $\delta$ should be negligible~\cite{papernot2018scalable}, particularly less than $\frac{1}{|D|}$. When $\delta=0$, we can call $(\epsilon,\delta)$-DP as $\epsilon$-DP.

Some new definitions have been derived from DP and applied in federated learning. For example, \cite{ghazi2021deep} defines label differential privacy (label DP) as follows.
It consider the situation where only labels are sensitive information that should be protected and is usually applied to differentially private vertical federated learning.

\vspace{-0.2cm}
\begin{definition}\label{def-label DP}
    ({\bf Label DP~\cite{ghazi2021deep}}). The randomized algorithm $\mathcal{A}:\mathbb{X}^n\to\mathbb{Y}$ satisfies $(\epsilon,\delta)$-label DP if any two neighboring datasets $D$ and $D'$ that differ in the label of a single sample and $D\simeq D^{\prime} \in \mathbb{X}^n$, we have 
    \begin{align*}
        \forall S\subseteq\mathbb{Y}:\mathrm{Pr}[\mathcal{A}(D)\in S]\leq e^{\epsilon}\mathrm{Pr}[\mathcal{A}(D^{\prime})\in S]+\delta.
    \end{align*}
\end{definition}
\vspace{-0.2cm}

On the other hand, Bayesian Differential Privacy (BDP)~\cite{triastcyn2020bayesian} is interested in the change in the posterior distribution of the attacker after observing the private model, compared to the prior. The original definition of BDP is very similar to DP, except that BDP assumes that all samples in the dataset are drawn from the same distribution $p(x)$. 

\vspace{-0.2cm}
\begin{definition}({\bf Bayesian DP \cite{triastcyn2020bayesian}})
The randomized algorithm $\mathcal{A}:\mathbb{X}^n\to\mathbb{Y}$ satisfies $(\epsilon,\delta)$-BDP if any two neighboring datasets $D$ and $D'$ that differ in a single data sample $x^{\prime}\sim p(x)$ 
where $p(x)$ represents specific probability distribution, we have 
    \begin{align*}
        \forall S\subseteq\mathbb{Y}:\mathrm{Pr}[\mathcal{A}(D)\in S]\leq e^{\epsilon}\mathrm{Pr}[\mathcal{A}(D^{\prime})\in S]+\delta.
    \end{align*}
\end{definition}
\vspace{-0.2cm}





\vspace{-0.2cm}
\subsubsection{LDP} 
Subsequent research introduced the concept of local differential privacy (LDP) framework, which is more stringent compared to DP. The difference between LDP and DP is that LDP does not require truthful data collectors. Therefore, data perturbation’s function is transferred from data collectors to each user. Each user perturbs original data by privacy-preserving algorithms and then uploads the disturbance data to data collectors. The formal definition of LDP is presented as follows.


\vspace{-0.2cm}
\begin{definition}\label{def-Local Differential Privacy}
    ({\bf LDP~\cite{evfimievski2003limiting}}). The randomized mechanism ${{\mathcal{R}}:\mathbb{X}\rightarrow\mathbb{Y}}$ satisfies $(\epsilon,\delta)$-LDP if for any two inputs $x,x^{\prime} \in \mathbb{X}$, we have
    \begin{align*}
        \forall t \in \mathbb{Y}:{\mathrm{Pr}[\mathcal{R}(x)=t]\leq e^{\epsilon}\operatorname*{Pr}[\mathcal{R}(x^{\prime})=t]}+\delta.
    \end{align*}
\end{definition}
\vspace{-0.2cm}

Just like the DP, $0 \leq \delta \leq 1$ indicates the failure probability that the property does not hold, and it should be negligible. And when $\delta=0$, we can call $(\epsilon,\delta)$-LDP as $\epsilon$-LDP. The definition we provided here is the general version, but the vast majority of works based on LDP utilize the definition of $\epsilon$-LDP.

From Definition~\ref{def-Local Differential Privacy}, it can be seen that LDP ensures that algorithm $\mathcal{R}$ satisfies $(\epsilon,\delta)$-LDP by controlling the similarity of any two records’ output. In short, according to a certain output result of privacy algorithm $\mathcal{A}$, it is almost impossible to infer which record its input data is. In DP, its privacy algorithm $\mathcal{A}$ is defined by neighbor dataset, so it requires truthful third-party data collectors to protect data analysis results. 
We can see that the main difference between DP and LDP lies in the definition of neighboring datasets. LDP requires that all pairs of sensitive data must satisfy the same $\epsilon$ privacy guarantee, which may hide too much information from the datasets, leading to insufficient utility for certain applications.
Gursoy et al.~\cite{gursoy2019secure} extended metric-based extensions of differetial privacy to LDP and proposed condensed LDP (CLDP) to improve the utility, which measures the level of privacy guarantee between any pair of sensitive data based on their distance.

\vspace{-0.2cm}
\begin{definition}\label{def-CLDP}
    ({\bf Condensed LDP~\cite{gursoy2019secure}}). The randomized mechanism ${{\mathcal{R}}:\mathbb{X}\rightarrow\mathbb{Y}}$ satisfies $\epsilon$-CLDP if for any two inputs $x,x^{\prime} \in \mathbb{X}$, we have
    \begin{align*}
        \forall t \in \mathbb{Y}:{\mathrm{Pr}[\mathcal{R}(x)=t]\leq e^{\epsilon \cdot d(x,x^{\prime})}\operatorname*{Pr}[\mathcal{R}(x^{\prime})=t]},
    \end{align*}
    \vspace{-0.2cm}
    where $d: \mathbb{X}\times\mathbb{X}\to[0,\infty)$ is a distance function that takes as input two items $x,x^{\prime} \in \mathbb{X}$.
\end{definition}

In $\epsilon$-CLDP,  the control of indistinguishability is influenced not only by $\epsilon$ but also by the input distance $d(\cdot,\cdot)$. Consequently, as $d$ increases, $\epsilon$ must decrease to maintain compensation.

\vspace{-0.2cm}
\subsubsection{Shuffle Model}
The shuffle differential privacy model, built upon the foundation of LDP. It adds a trusted shuffler between the server and the clients, which shuffles the data items submitted by clients to achieve anonymization. In shuffle model, each client satisfies privacy guarantee of LDP when facing shuffler, and then achieves privacy guarantee of DP when facing the server since the shuffled data forms a dataset and generates the notion of neighborhood dataset.

The shuffle model originated from the framework of \emph{Encode, Shuffle, Analyze} proposed by Bittau et al.~\cite{bittau2017prochlo} and is formally defined by Cheu et al. \cite{cheu2019distributed}. Following these works, a protocol in the shuffle model consists of three components $\mathcal{P}={\mathcal{A}\circ\mathcal{S}\circ\mathcal{R}^{n}}$, aiming to achieve $(\epsilon_c,\delta)$-DP. Here, $\mathcal{R}:\mathbb{X}\rightarrow\mathbb{Y}$ is a local randomizer executed on the client side, usually assumed to satisfy $\epsilon_l$-LDP. ${\mathcal{S}:\mathbb{Y}^{n}\to\mathbb{Y}^{n}}$ is a shuffler that applies a random permutation to its inputs, achieving anonymity. ${\mathcal{A}:\mathbb{Y}^{n}}\to\mathbb{Z}$ is the analyzer, aggregating the received values for statistics. According to the post-processing invariance, if the protocol $\mathcal{M} = \mathcal{S}\circ\mathcal{R}^{n}$ satisfies $(\epsilon_c,\delta)$-DP, the whole protocol $P$ also satisfies $(\epsilon_c,\delta)$-DP. Thanks to the anonymity brought by shuffling, each user's privacy can be not only by their local randomization but also the randomness provided by other clients, leading to a smaller centralized privacy loss (i.e., $\epsilon_c$) compared to the used local privacy budget (i.e., $\epsilon_l$), just like the Figure~\ref{figure: shuffle model} shown. In theory, we call this phenomenon privacy amplification, which leads to an around $O(\sqrt{n})$ more minor privacy loss $\epsilon_c$ \cite{balle2019privacy, erlingsson2019amplification}. This also reflects that to achieve a similar privacy level (corresponding to the same $\epsilon$), less noise is required for the local randomizer.

\vspace{-0.2cm}
\subsubsection{Relation between DP, LDP and shuffle model}\label{subsec:Relation between DP, LDP and shuffle}
In this section, we will discuss the relationships and differences between the DP, LDP, and shuffle model.

\begin{figure*}[h]
  \centering
  \begin{subfigure}{0.45\linewidth}
    \centering
	 \includegraphics[width=0.8\linewidth]{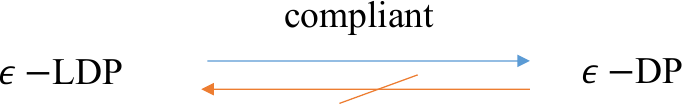}
		\caption{relation between LDP and DP.}
		\label{figure:relation between LDP and DP}
  \end{subfigure}
  \hfill
  \begin{subfigure}{0.50\linewidth}
    \centering
    \includegraphics[width=0.9\linewidth]{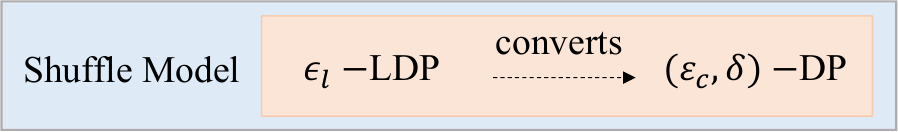}
		\caption{Shuffle model, DP and LDP.}
		\label{figure:shuffle model, DP and LDP}
  \end{subfigure}
  \caption{Relation between DP, LDP and shuffle model.}
  \label{figure: DP LDP and Shuffle Model}
  \vspace{-0.3cm}
\end{figure*}

\textbf{DP and LDP.} We can see that the main difference between DP and LDP lies in their definitions: DP relies on neighboring datasets as input, while LDP does not have the concept of neighboring datasets. Alternatively, from another perspective, when there is only one data sample, DP and LDP are equivalent. So, the LDP~(definition~ \ref{def-Local Differential Privacy}) can be derived from the DP~(definition~\ref{def-Differential Privacy}) when the input $x$, $x^{\prime}$ are taken to be datasets of only one record. Since the size of local dataset is 1, $x$ and $x^{\prime}$ are neighbors for all $x,x^{\prime} \in \mathbb{X}$. Therefore LDP is a stronger condition as it requires the mechanism to satisfy DP for any two values of the domain of data $X$. As shown in Figure~\ref{figure:relation between LDP and DP}, LDP can compliant the DP, but the opposite doesn't hold true~\cite{paul2020ara,chen2023differential}.

\textbf{Shuffle model, DP and LDP.}
The shuffle model encompasses both LDP and DP techniques and concepts. As shown in Figure~\ref{figure:shuffle model, DP and LDP}, it begins with the definition of LDP, where each client satisfies $\epsilon_l$-LDP protection, and then sends to the server using the shuffle model. At this point, the shuffle model achieves anonymity of model parameters and forms a dataset, thereby generating the concept of neighborhood datasets. When facing the server, all uploaded model collections will be made to satisfy $(\epsilon_c,
\delta_c)$-DP. As mentioned earlier, while LDP can compliant the DP, the shuffle model can achieve a tighter privacy budget bound in DP, meaning that $\epsilon_c \ll \epsilon_l$.

\vspace{-0.2cm}
\section{Differentially Private HFL} \label{sec-Horizontal}
Differentially private federated learning research has largely focused on the HFL scenario. Therefore, in this section, we will specifically introduce the implementation of various DP models within HFL. Specifically, we will show the works of differentially private HFL from DP, LDP and shuffle model separately.

\begin{figure*}[h]
  \centering
  \begin{subfigure}{0.55\linewidth}
    \centering
    \includegraphics[width=1.0\linewidth]{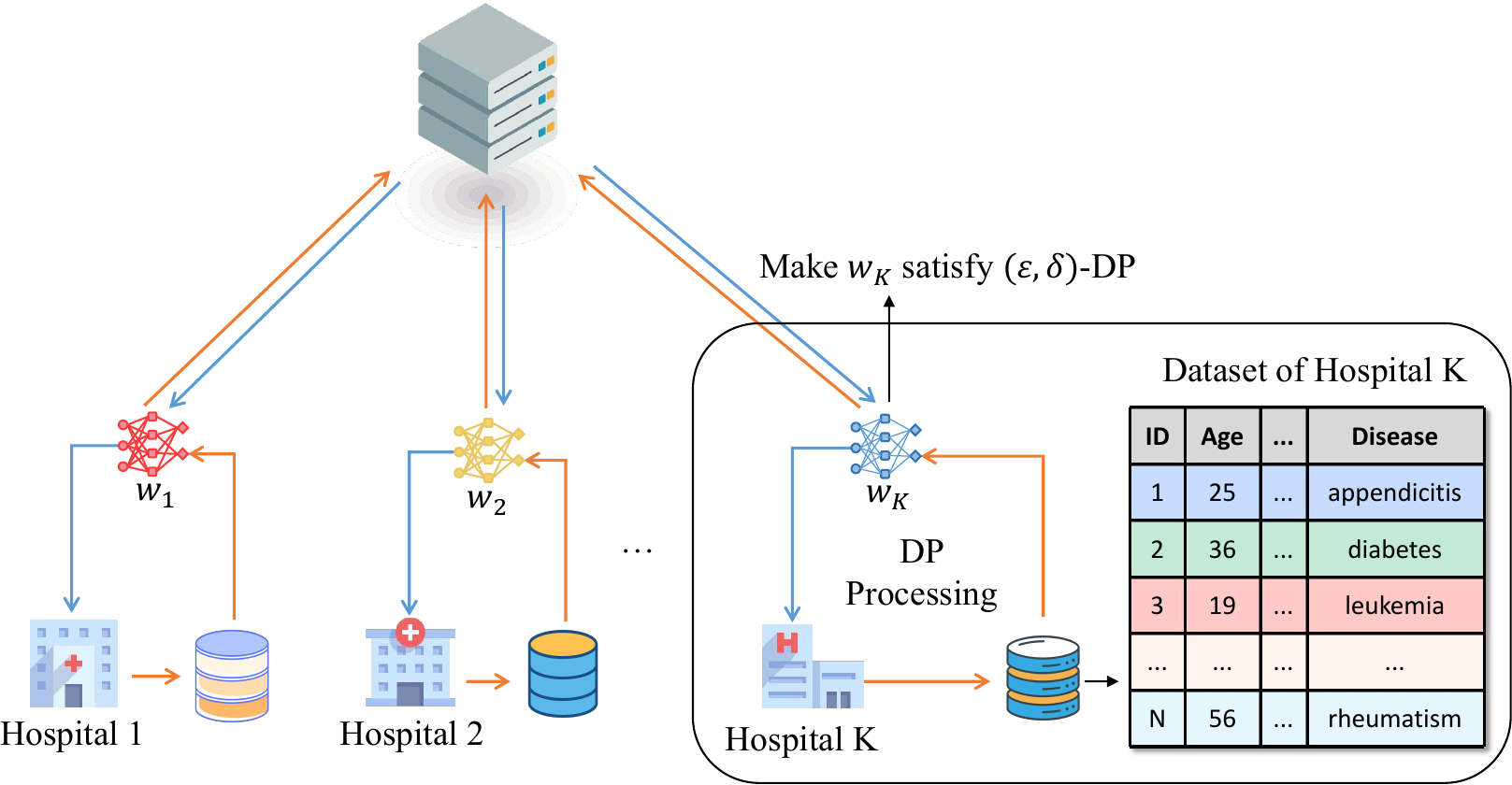}
    \caption{SL-DP.}
    \label{figure: SL-DP}
  \end{subfigure}
  \hfill
  \begin{subfigure}{0.44\linewidth}
    \centering
    \includegraphics[width=1.0\linewidth]{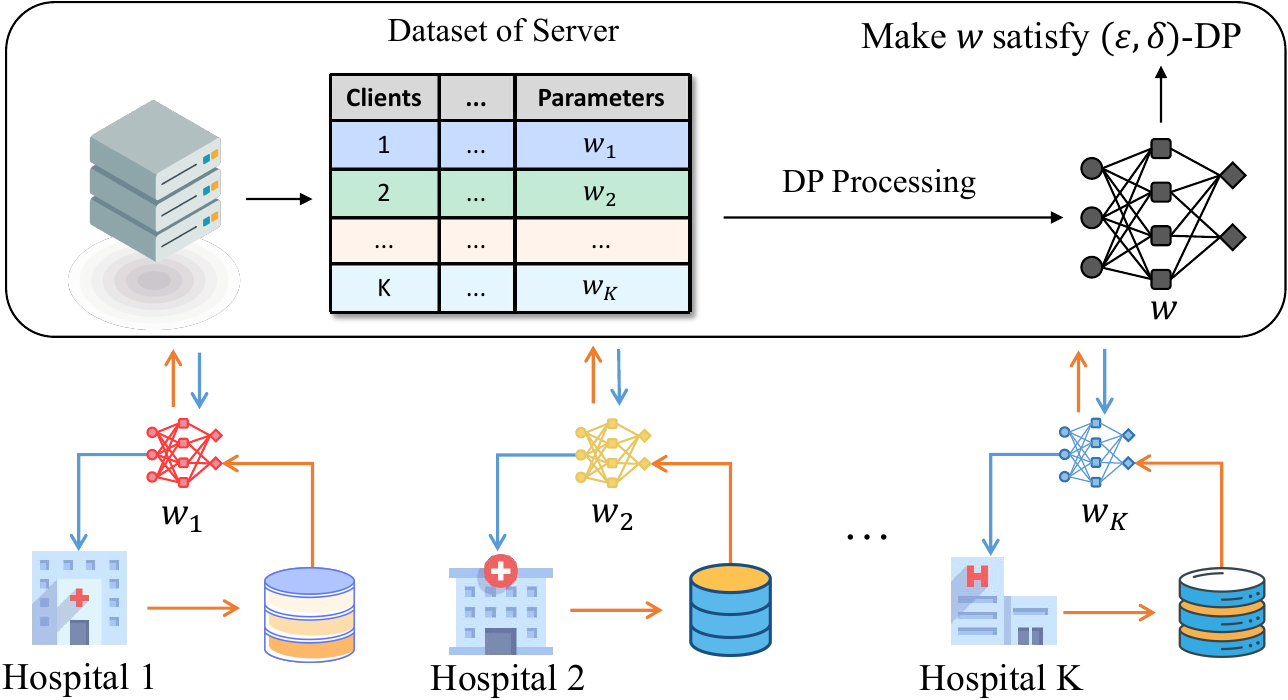}
    \caption{CL-DP.}
    \label{figure: CL-DP}
  \end{subfigure}
  \caption{SL-DP and CL-DP in HFL.}
  \vspace{-0.4cm}
  \label{figure: SL-DP and CL-DP}
\end{figure*}

\vspace{-0.2cm}
\subsection{DP-HFL} \label{subsec-cdp}
As mentioned in Section~\ref{subsec-dp}, the concept of DP is based on the input of two neighboring datasets. The definition of neighbouring datasets on federated learning depends on the desired formulation of privacy in the setting (i.e. which objects need to be kept private). 

As in Figure~\ref{figure: SL-DP and CL-DP}, in general HFL, we can find two data owners, the local client and the central server. Therefore, based on the notions of neighboring datasets, the two main levels of neighboring in HFL can to be formally defined as follows: 

\vspace{-0.2cm}
\begin{definition} \label{def-sl}
    ({\bf Sample-level DP (SL-DP)}). Under SL-DP, two datasets $D$ and $D^{\prime}$ are neighbouring if they
differ in a single sample or record (either through addition or through removal).
\end{definition}
\vspace{-0.2cm}

\vspace{-0.2cm}
\begin{definition} \label{def-cl}
    ({\bf Client-level DP (CL-DP)}\footnote{Some articles treat ``user-level DP'' as equivalent to ``client- level DP'' because they assume that each client has only one user~\cite{cheng2022differentially,yang2023dynamic}. However, we provide a more general definition, where we consider the possibility of one or more users under a single client. Further discussion and details regarding the definition and functionality of user-level are addressed in Section~\ref{sec-future direction}.}). Under CL-DP, two datasets $D$ and $D^{\prime}$ are neighbouring if they
differ in a single client or device (either through addition or through removal).
\end{definition}
\vspace{-0.2cm}

\vspace{-0.2cm}
\subsubsection{SL-DP}  \label{subsubsec-sl}
Under SL-DP, the data owner can be each local client. As shown in the Figure~\ref{figure: SL-DP}, each hospital participating in federated learning has a local database. For each client, the goal of SL-DP is to hide the presence of a single sample or record, or to be more specific, to bound the influence of any single sample or record on the learning outcome distribution (i.e. the distribution of the model parameters). It protects each local sample or record, so that the attackers could not identify one sample from the union of all local datasets.

Generally, at the SL-DP, we assume that the server is semi-honest (honest but curious). Therefore, each client cares about the privacy of their local dataset, and each client's privacy budget is independent. The most mainstream approach to implementing SL-DP is to execute the Differentially Private Stochastic Gradient Descent (DPSGD) algorithm~\cite{abadi2016deep} on each client~\footnote{If one wishes to define SL-DP under the aggregate samples held by all clients in FL, it requires the establishment of a secure third party. For example, Ruan et al.~\cite{ruan2023private} proposed a secure DPSGD algorithm by combining differential privacy and secure multiparty computation. They designed a secure inverse of square root method to securely clipping the gradient vectors and a secure Gaussian noise generation protocol. Their algorithm can efficiently perform DPSGD in ciphertext.}. DPSGD is a widely-adopted training algorithm for deep neural networks with differential privacy guarantees. Specifically, in each iteration $t$, a batch of tuples $\mathcal{B}_t$ is sampled from $D$ with a fixed probability $\frac{b}{|D|}$, where $b$ is the batch size. After computing the gradient of each tuple $x_i \in \mathcal{B}_t$ as $g_t(x_i) = \nabla_{\theta_i} L(\theta_i,x_i)$, where $\theta_i$ is model parameter for the i-th sample, DPSGD clips each per-sample gradient according to a fixed $\ell_{2}$ norm (Equation~\eqref{eq:clipping}).
\vspace{-0.1cm}
\begin{align}
	\begin{split}
		\overline{g}_t\left(x_{i}\right)
		& = \textbf{Clip}(g_t\left(x_{i}\right);C) \label{eq:clipping} = g_t\left(x_{i}\right) \Big/ \max \Big(1, 		\frac{\left\|g_t\left(x_{i}\right)\right\|_{2}}{C}\Big).
	\end{split}
\end{align}

In this way, for any two neighboring datasets, the sensitivity of the query $\sum_{i\in \mathcal{B}_t} g(x_i)$ is bounded by $C$. Then, it adds Gaussian noise scaling with $C$ to the sum of the gradients when computing the batch-averaged gradients:
\begin{equation}\label{eq:add noise}
\tilde{g}_t = \frac{1}{b}\left(\sum_{i \in \mathcal{B}_t} \overline{g}_t\left(x_{i}\right)+\mathcal{N}\left(0, \sigma^{2} C^{2} \mathbf{I}\right)\right),
\end{equation}
where $\sigma$ is the noise multiplier depending on the privacy budget. Last, the gradient descent is performed based on the batch-averaged gradients. Since initial models are randomly generated and independent of the sample data, and the batch-averaged gradients satisfy the differential privacy, the resulted models also satisfy the differential privacy due to the post-processing property. Many state-of-the-art variants of DPSGD~\cite{fu2023dpsur,wei2022dpis,papernot2021tempered} can also be directly applied in the local client iterations to achieve more efficient model under SL-DP.


However, the impact of data heterogeneity is a key obstacle in DP-FL. Huang et al.~\cite{huang2020dp} proposed differential privacy convolutional neural network with adaptive gradient descent algorithm (DPAGD-CNN) to update the training parameters of each client. They selected the best learning rate from a
candidate set based on model evaluation in each round of local DPSGD. Noble et al.~\cite{noble2022differentially} introduced DP-SCAFFLOD, an extension of the SCAFFLOD algorithm~\cite{karimireddy2020scaffold}, which incorporates differential privacy. It employs a control variable to limit model drift during local DPSGD, aligning them more closely with the global model direction. Wei et al.~\cite{wei2021user} found that there is an optimal number of communication rounds in terms of convergence performance for a given privacy budget $\epsilon$, which has motivated them to adaptively allocate privacy budgets in each round. Fu et al.~\cite{fu2022adap} proposed the Adap DP-FL algorithm, which includes adaptive gradient clipping and adaptive noise scale reduction methods. In the gradient clipping step of DPSGD, gradients are clipped using adaptive thresholds to account for the heterogeneity of gradient magnitudes across clients and training rounds. Additionally, during noise addition, the noise scale gradually decreases as gradients converge across training rounds. Yang et al.~\cite{yang2023privatefl} start from the Non-IID data itself, by concurrently updating local data transformation layers during local model training, thereby reducing the additional heterogeneity introduced by DP, and consequently improving the utility of FL models for Non-IID data. Their method was also applied to the CL-DP setting. Chen et al.~\cite{chen2024differentially} utilized a adaptive server-side optimization method, simultaneously mitigating the negative impact of non-IID data and DP noise on model utility.

The allocation of privacy budget is also the focus of the study, as the privacy budget determines factors such as the number of iteration rounds and noise scale, it directly impacts the performance of the trained model. Ling et al.~\cite{ling2024ali} investigated how to achieve better model performance under constraints of privacy budget and communication resources. They conducted convergence analysis of DP-HFL, derived the optimal number of local iterations before each aggregation. Liu et al.~\cite{liu2021projected} considered scenarios with heterogeneous privacy budgets and proposed the Projected Federated Averaging (PFA) algorithm. This algorithm utilizes the top singular subspace of model updates submitted by clients with higher privacy budgets to project them onto model updates from clients with lower privacy budgets. Furthermore, Liu et al.~\cite{liu2024cross} pioneered the formulation of sample-level personalized differential privacy in federated learning. They introduced a two-stage hybrid sampling scheme to satisfy personalized privacy budget requirements at the sample level and provided a theoretical analysis of the privacy amplification effect for the mechanism.
 In addition to heterogeneous privacy budgets in each client, malekmohammadi et al.~\cite{malekmohammadi2024noise} also considered the heterogeneity of client batching and dataset sizes to propose Robust-HDP, which effectively estimates the true noise level in client model updates and substantially reduces the noise in aggregated model updates. Furthermore, it reduces communication overhead by having clients with lower privacy budgets upload projected model updates instead of original model values. Zheng et al.~\cite{zheng2021federated} integrated GDP privacy metric into DP-HFL, proposing a private federated learning framework called PriFedSync. While considering the cost of communication, Li et al.~\cite{li2022soteriafl} proposed SoteriaFL, which is a unified framework for compressed private FL. They use shifted compression scheme~\cite{horvath2023stochastic} to compress the perturbed parameters for efficient communication while maintaining high utility. 

 In addition to this, related articles have explored more complex attack scenarios under SL-DP protection. Xiang et al.~\cite{xiang2023practical}  proposed a method that combines local DPSGD with Byzantine fault tolerance techniques. By leveraging differential privacy's random noise, they construct an aggregation approach that effectively thwarts many existing Byzantine attacks. This method ensures the privacy and stability of federated learning systems by introducing randomness that prevents Byzantine attackers from accurately interfering with gradient updates and aggregation processes. Wei et al.~\cite{wei2020federated} assumed that downlink broadcast channels are more dangerous than uplink channels and proposed NBAFL, which not only add noise to the parameters before aggregation but also add noise again after aggregation to achieve a higher level of privacy protection. Naseri et al.~\cite{naseri2020local} proposed a analytical framework that empirically assesses the feasibility and effectiveness of SL-DP and CL-DP in protecting FL. Through many attack experiments, their results indicate that while SL-DP can defend membership inference attacks and backdoor attacks, it cannot resist attribute inference attacks. 

In addition to the above stochastic gradient descent, alternating direction method of multipliers (ADMM) is also local optimization method for FL. ADMM introduces Lagrange multipliers to transform the original problem into a series of subproblems, which are then alternately solved until convergence~\cite{zhang2014asynchronous}.
Huang et al.~\cite{huang2019dp} utilize a first-order approximation function as the objective function for local clients. This first-order approximation function is convex and naturally constrains the $l_2$ norm of the gradient within a bound without clipping the gradient to get the sensitivity. Additionally, they devised an adaptive noise coefficient decay to facilitate the convergence of this first-order approximation function. Based on this, Ryu et al.~\cite{ryu2022differentially} proposed local multiple iterations to accelerate convergence speed and reduce privacy loss. However, ADMM-based optimization methods can only be used for convex optimization problems and are not widely applicable in FL.

\vspace{-0.2cm}
\subsubsection{CL-DP}   \label{subsubsec-cl}
Under general CL-DP, the server is often assumed to be completely honest (honest and not curious) and can be viewed a data owner. As shown in the figure~\ref{figure: CL-DP}, the central server can collect the parameters uploaded by each client in each round. The goal of CL-DP is to hide the presence of a single client or device, or to be more specific, to limit the impact of any single client or device on the distribution of aggregation results. It requires that the attackers cannot identify the participation
of one client or device by observing the output of aggregated parameters~\footnote{It is worth noting that while some works locally clip and add noise, the amount of noise added is $\mathcal{N}(0,\sigma^2C^2/K)$, where $K$ is the number of clients participating in federated learning~\cite{shi2023make,cheng2022differentially}. Each client model $\mathbf{w}_{k}$ satisfy weak LDP protection, but this is not the protection goal. The intention is for the uploaded data to the central aggregator to be equivalent to $\sum_{k=1}^K \mathbf{w}_{k} + \mathcal{N}(0,\mathbf{\sigma}^2C^2)$, it can be seen that the definition of neighboring datasets is CL-DP (Definition~\ref{def-cl}).}. 


Geyer et al.~\cite{geyer2017differentially} introduced CL-DP in federated learning for the first time, assuming differential attacks from any participating client. The server perturbs the distribution of the summed parameters by limiting each client's parameter contribution and then adding noise. Based on work of Geyer et al.~\cite{geyer2017differentially}, Mcmahan et al.~\cite{mcmahan2017learning} proposed the DP-FedAvg and DP-FedSGD algorithms, which employ client sampling techniques for privacy amplification and utilize the moment accountant mechanism~\cite{abadi2016deep} for privacy computation. Their algorithms achieved comparable performance of non-private models in Long Short-Term Memory (LSTM) language.

The previous work constrained each client's model parameters to a fixed constant, while Andrew et al.~\cite{andrew2021differentially} proposed an adaptive clipping method. Their method uses an adaptive clipping bound based on a parameter's norm distribution quantile, estimated with differential privacy, rather than a fixed bound.
Their experiments demonstrate that adaptive clipping to the median update norm performs well across a range of federated learning tasks. Zhang et al.~\cite{zhang2022understanding} studied the impact of clipping on model parameters and gradients. They demonstrated that uploading clipped gradients leads a better performance than uploading clipped models on the client side, and provided a convergence analysis based on gradient clipping with noise, where the upper bound of convergence highlighted the additional terms introduced by differential privacy.

In addition to this, many studies have focused on reducing the impact of noise on uploaded model parameters. Chen et al.~\cite{cheng2022differentially} observed that using a small clipping threshold can decrease the injected noise volume. They reduced the norm of local updates by regularizing local
models and making the local updates sparse. DP-FedSAM ~\cite{shi2023make} uses a SAM optimizer~\cite{foret2020sharpness} to help model parameters escape saddle points and enhance the robustness of the model to noise, aiming to find more stable convergence points. Bietti et al.~\cite{bietti2022personalization} proposed PPSGD, which trains personalized local models and ensures global model privacy simultaneously. They utilized personalized models to enhance the global model's performance. Similarly, Yang et al.~\cite{yang2023dynamic} dynamically preserved high-information model parameters locally from noise impact using layer-wise Fisher information. They also introduced an adaptive regularization strategy to impose differential constraints on model parameters uploaded to the server, enhancing robustness to clipping. Xu et al.~\cite{xu2023learning} took into account that the size of the softmax layer of model parameters is linearly related to the number of sample labels. By keeping the softmax layer local and not uploading it, the participation of more clients (i.e., more sample labels) in federated learning does not result in more noise injection. Triastcyn et al.~\cite{triastcyn2019federated} leveraged the assumption that data is distributed similarly across clients in HFL, which leads to their updates being highly consistent, and utilized BDP for privacy auditing to obtain tighter privacy bounds. Their assumption and method is not only used under the CL-DP, but also applied in SL-DP scenario.

\begin{figure}[htb]
	\begin{center}
\includegraphics[width=0.55\linewidth]{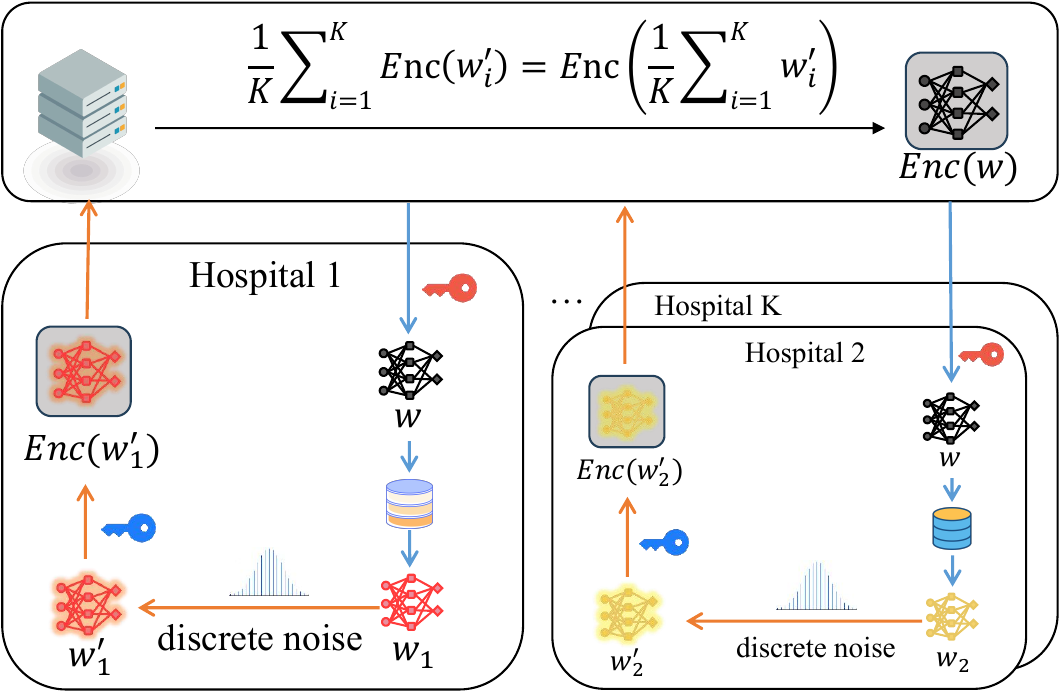}
		\caption{CL-DP with SA.}
		\label{fig: CL-DP with SA.}
	\end{center}
\vspace{-0.4cm}
\end{figure}

\textbf{CL-DP with Secure Aggregation (SA)~\footnote{Many works refer to this framework as ``distributed dp''~\cite{kairouz2021distributed,agarwal2021skellam,yang2023privatefl}.}.} As mentioned above, CL-DP often requires a secure server for aggregation. In order to eliminate the reliance on a trusted central authority, many recent studies have combined differential privacy techniques with secure aggregation~\cite{bonawitz2017practical} to achieve CL-DP. As shown in Figure~\ref{fig: CL-DP with SA.}, after local training, each client adds a small amount of differential privacy noise to the model parameters, then encrypts and sends them to the server. The server performs aggregation, enabling the aggregation of noise levels from each client, thereby ensuring that the encrypted model satisfies CL-DP. However, secure aggregation relies on modular arithmetic and cannot be directly compatible with continuous noise mechanisms (such as Laplace mechanism and Gaussian mechanism) currently used to implement differential privacy. This has led researchers to begin designing discrete noise addition mechanisms and integrating them with secure aggregation to characterize differential privacy.

Agarwal et al.~\cite{agarwal2018cpsgd} proposed and extended the first discrete mechanism, namely the binomial mechanism, in the federated setting. The discrete nature of binomial noise, along with quantization techniques~\cite{suresh2017distributed}, enables efficient transmission. However, applying this mechanism to practical tasks faces limitations. Firstly, binomial noise can only achieve approximate differential privacy with nonzero probability of fully exposing private data, and the binomial mechanism does not achieve Renyi or concentrated differential privacy.

To address these issues, Wang et al.~\cite{wang2020d2p} first discretized local parameters~\cite{mcmahan2017communication}, then added noise from the Discrete Gaussian Distribution to satisfy differential privacy. Kairouz et al.~\cite{kairouz2021distributed} also employed a Discrete Gaussian Mechanism, providing a novel privacy analysis associated with the proportion of malicious clients for the sum of Discrete Gaussian Mechanism. They extensively investigate the impact of discretization, noise, and modular clipping on model utility. However, the Discrete Gaussian Mechanism suffers from the problem of non-closure under addition. To address this, Agarwal et al.~\cite{agarwal2021skellam} proposed the Skellam Mechanism, which uses the difference of two independent Poisson distributed random variables as noise. The noise distribution of this mechanism naturally satisfies the property of closure under addition due to the first-kind modified Bessel function, thereby avoiding additional privacy budget for noise summation. However, the noise magnitude of the aforementioned discrete noise schemes is unbounded, thus requiring modular clipping for secure aggregation, leading to additional bias. Chen et al.~\cite{chen2022poisson}introduced the multi-dimensional Poisson Binomial Mechanism, a unbiased and bounded discrete differential privacy mechanism. It treats model parameters as probabilities for a binomial distribution, and generates binomial noise based on these probabilities to control the noise bound. 

However, the above methods rely on per-parameter quantization, resulting in significant communication overheads. To further explore the fundamental communication costs required for achieving optimal accuracy under centralized differential privacy, Chen et al.~\cite{chen2022fundamental} proposed using a linear scheme based on sparse random projection to reduce communication overheads. They utilize count sketch matrices and hash functions to obtain sparse projection matrices, thus reducing model parameter dimensionality. Kerkouche et al.~\cite{kerkouche2021compression} also sparsify vectors by decomposing model parameters into a ``sparse orthogonal basis matrix'' and a ``sparse signal'' multiplied by the discrete Gaussian mechanism locally. The server can reconstruct the sparse gradient vector by solving a convex quadratic optimization problem.  In addition to designing new discrete noise mechanisms to integrate secure aggregation techniques, Stevens et al.~\cite{stevens2022efficient} proposed a secure federated aggregation scheme based on the Learning With Errors (LWE) problem~\cite{regev2009lattices}. It encrypts the model parameters of local clients using masks generated by the LWE problem before uploading. This method utilizes the random noise naturally introduced by LWE technology, ensuring that the sum of errors in the LWE problem satisfies differential privacy. Kato et al.~\cite{kato2023olive} implement CL-DP by constructing a trusted aggregator using Trusted Execution Environment (TEE).


\vspace{-0.2cm}
\subsection{LDP-HFL} \label{subsec-ldp}
LDP-HFL can be viewed as a mean estimation problem based on LDP, as each model parameter represents a high-dimensional continuous data point. Each client perturbs their model parameters locally before sending them to the server, which aggregates the data and produces the mean estimation result. There has been extensive research on mean estimation based on LDP~\cite{duchi2013local,nguyen2016collecting}, but directly applying it to federated learning faces certain obstacles. This is because the dimensionality of model parameters in federated learning is extremely high, and as the dimension increases, the privacy budget allocated to each dimension decreases, leading to an exponential increase in statistical variance~\cite{duchi2018minimax}.

Many scholars are studying new mean estimation techniques of high-dimensional data under LDP. Wang et al.~\cite{wang2019collecting} proposed Piecewise Mechanism (PM) and Hybrid
Mechanism (HM), which can perturb multidimensional data containing both numeric and categorical data under optimal worst-case error. Furthermore, they presented an LDP-compliant algorithm for FL using PM and HM, and got the high utility. Zhao et al.~\cite{zhao2020local} proposed two new LDP mechanisms. Building upon Duchi~\cite{duchi2013local}, they introduced the Three-Outputs mechanism, which offers three discrete output possibilities. This mechanism achieves a small worst-case noise variance under small privacy budget $\epsilon$. For larger privacy budget $\epsilon$, inspired by the multiple-outputs strategy, they devised an optimal partitioning mechanism based on the PM~\cite{wang2019collecting}, named PM-SUB. Trues et al.~\cite{truex2020ldp} converted the each element of the local parameters to a discrete space, and then perturbed them separately by exponential mechanism under Condensed-LDP~\cite{gursoy2019secure}.  They applied these two mechanisms to the parameters uploaded from clients to the server to achieve LDP-FedSGD.

Building on the optimization of existing LDP mechanisms, many studies utilize parameters shuffling to further mitigate the curse of dimensionality in LDP-HFL. Sun et al.~\cite{sun2021ldp} proposed the Adaptive-Duchi mechanism, which sets an adaptive perturbation range for each layer of parameters in the model base on Duchi~\cite{duchi2018minimax}. Furthermore, the parameter shuffling method is proposed by them, where they split the model parameters of each client by layer and then shuffle users' parameters within each layer. If regarding the information in different layers as independent in some scenarios, the privacy budget will no longer need to be split according to the parallel property of DP, further improving utility~\footnote{
Although this method involves the action of shuffling, it does not conform to the concept and definition of the shuffle model. Parameter shuffling only avoids the division of the privacy budget across multiple dimensions under LDP based on the assumption that parameters are independent at each layer, but it does not achieve privacy amplification from LDP to DP.}. Zhao et al.~\cite{zhao2022privacy} also employed parameter shuffling to eliminate the associations between dimensions. In addition, they designed an Adaptive-Harmony mechanism to allocate perturbation interval adaptively to the parameters in each layer of the model. Varun et al.~\cite{Varun24SSRFL} proposed SRR-FL, in which they used the Staircase Randomized Response (SRR) mechanism~\cite{wang2022srr} for local perturbations before parameter shuffling. The SRR mechanism differentiates by distributing different perturbation probabilities among various value groups within the domain, with higher perturbation probabilities assigned to values closer to the true value. This approach can enhance the performance and utility of the randomization scheme.

The other method to alleviate the curse of dimensionality is select or sample dimension . 
Liu et al.~\cite{liu2020fedsel} proposed FedSel, which is a two-step LDP-FL framework that includes a dimension selection stage and a value perturbation stage. In dimension selection step, they build a top-$k$ dimension set containing the
dimensions of the $k$ largest absolute update values and privately selects one ``importan'' dimension
from the top-$k$ set. Then the value of selected dimension is perturbed by Peicewise Mechanism~\cite{wang2019collecting} in stage two.
However, FedSel only selects one dimension for each local model parameter, which may lead to slower model convergence. Jiang et al.~\cite{jiang2022signds} extended it to multi-dimensional selection and designed the Exponential Mechanism-based Multi-Dimension Selection (EM-MDS) algorithm to reduce the privacy budget incurred when selecting multiple dimensions. Additionally, they assigned a sign variable to the selected dimension values instead of directly perturbing them. To address the slow convergence caused by dimension selection, Li et al.~\cite{li2022fedta} proposed iterating locally using the Adam optimizer~\cite{chollet2015keras}, then selecting the top-k dimensions and adding noise using the Laplace mechanism. Wang et al.~\cite{wang2023ppefl} aslo use exponential mechanism to choose Top-$K$ dimensions before perturbed raw parameters. In addition this, they proposed DMP-UE mechanism to perturb the Top-$K$ parameters get high utility, which extends upon Duchi~\cite{duchi2013local} to output three cases (including output 0), rather than two. They also reduced communication costs by introducing edge nodes to perform edge aggregation.

Other research focuses on the heterogeneity of clients in federated scenarios. For the data heterogeneity,
Zhang et al.~\cite{zhang2023pfldp} focused on the performance degradation caused by model heterogeneity in non-i.i.d. data settings, they proposed a personalized federated learning approach (FedBDP) based on Bregman divergence and differential privacy. Their algorithm employs Bregman divergence to quantify the discrepancy between local and global parameters and incorporates it as a regularization term in the update loss function. Additionally, by defining decay coefficients, they dynamically adjust the magnitude of the differential privacy noise for each round. Wang et al.~\cite{wang2020federated} introduced FedLDA, an LDP-based latent Dirichlet allocation (LDA) model tailored for federated learning settings. FedLDA employs an innovative random response mechanism with a prior (RRP), ensuring that the privacy budget remains independent of the dictionary size. This approach enhances accuracy through adaptive and non-uniform sampling processing.
In response to the heterogeneous device scenario, Lian et al.~\cite{lian2022webfed} proposed a browser-based cross-platform federated learning framework called WebFed. To enhance privacy protection, each client adds Laplace noise to the weights of their local models before uploading the training results. And for heterogeneity in privacy budgets, Yang et al.~\cite{yang2021federated} proposed PLU-FedOA, Which focus on various privacy requirements of clients. Firstly, they added Laplace noise to parameters locally using different privacy budget $\epsilon_k$ to achieve personalized LDP, and then designed a parameters aggregation algorithm for serve to close to the unbiased estimate. Zhang et al.~\cite{zhang2024dynamic} improved model performance by dynamically allocating the privacy budget at each federated aggregation round.

Last but not least, Some methods choose to perturb raw data directly rather than model parameters.
Wang et al.~\cite{wang2022safeguarding} treat each data point as an individual user and locally perturb each raw data before uploading it to the edge server. They employ the RAPPOR method~\cite{erlingsson2014rappor} to encode each feature of the data, followed by individual bit flips. Their experiments demonstrate that this local data perturbation approach effectively withstands data reconstruction attacks while maintaining model efficiency. Unlike directly perturbing the local raw data, Mahawaga et al.~\cite{mahawaga2022local} first extract feature vectors from the local datasets using the convolutional and pooling layers in the CNN model. These feature vectors are then converted into flattened 1-D vectors and subjected to unary encoding and perturbation by RAPPOR~\cite{erlingsson2014rappor} before being uploaded to the server. Wang et al.~\cite{wang2019sparse, wang2023generalized} analyse the theoretical error in generalized linear models under LDP.

\vspace{-0.2cm}
\subsection{Shuffle model-HFL} \label{subsec-Shuffle Modle}

The shuffle model combines LDP randomization with shuffling, ensuring the guarantee of DP against the analyzer. As such, referring to the taxonomy of HFL in DP, we divide HFL within the shuffle model into two classes, client level (i.e., CL-DP) and sample level (i.e., SL-DP), elaborated as follows.

\vspace{-0.2cm}

\subsubsection{Shuffle model of CL-DP}

\begin{figure*}[h]
  \centering
  \begin{subfigure}{0.44\linewidth}
    \centering
		\includegraphics[width=1.0\linewidth]{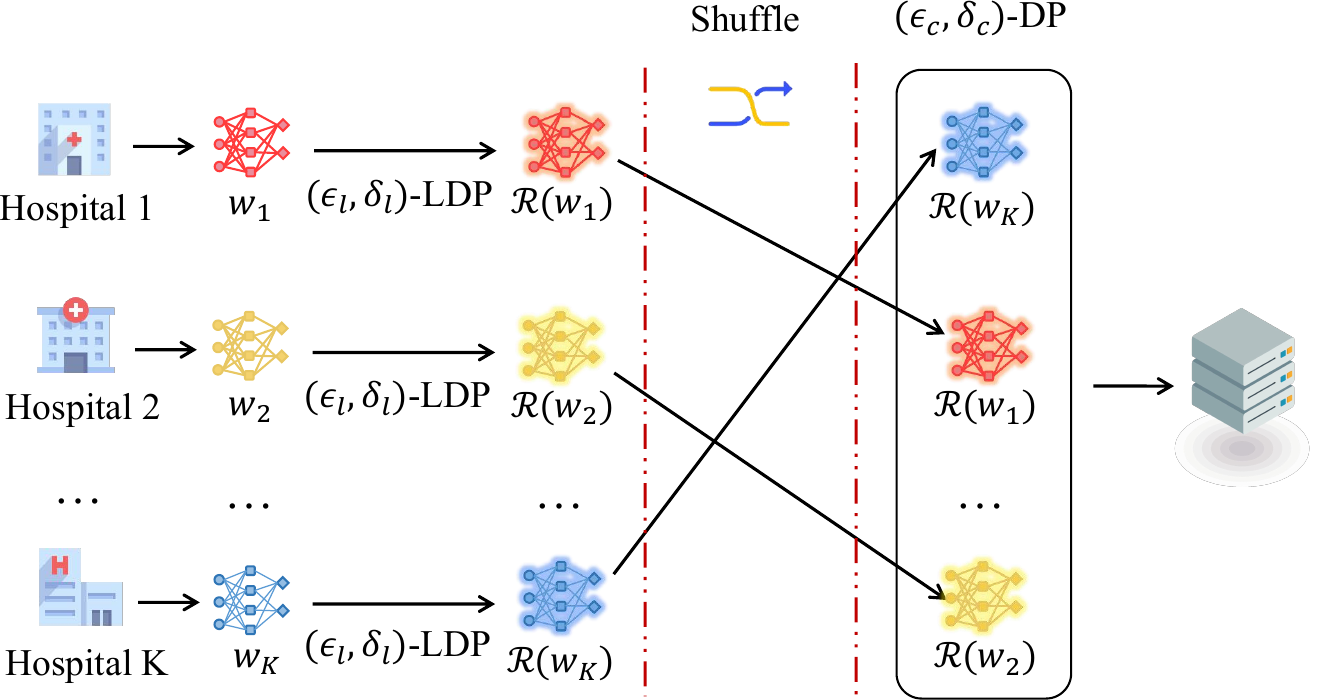}
		\caption{Shuffle model of CL-DP.}
		\label{figure:Shuffle1}
  \end{subfigure}
  \hfill
  \begin{subfigure}{0.55\linewidth}
    \centering
		\includegraphics[width=1.0\linewidth]{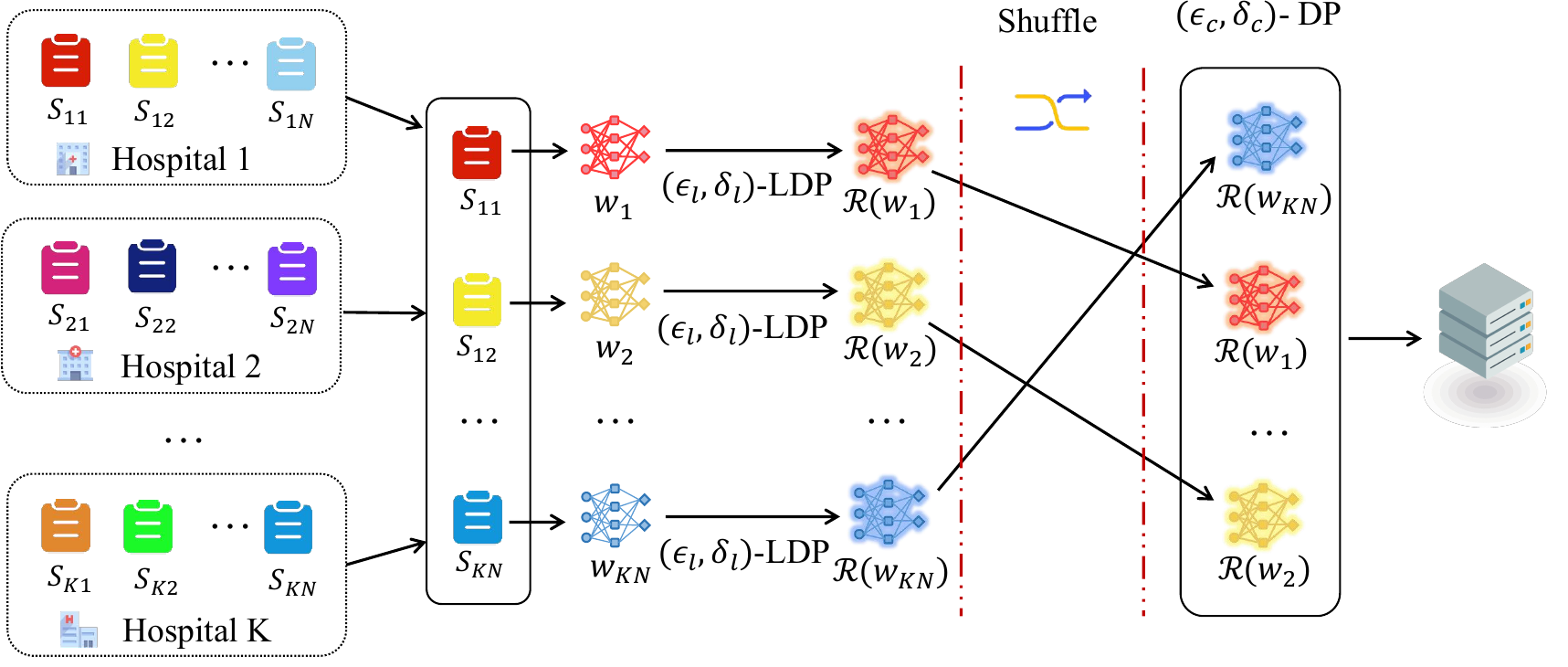}
		\caption{Shuffle model of SL-DP.}
		\label{figure:Shuffle2}
  \end{subfigure}
  \caption{Shuffle model in HFL.}
  \label{figure: shuffle model}
  \vspace{-0.4cm}
\end{figure*}

To achieve the shuffle model with CL-DP as shown in Figure~\ref{figure: shuffle model}~(a), each user randomizes their local updates with $(\epsilon_l, \delta_l)$-LDP, and the aggregated updates through shuffling satisfies $(\epsilon_c, \delta_c)$-DP. Typically, 
following the inertia of LDP, researchers consider $\epsilon_l$-LDP, setting $\delta_l = 0$. 
Note that we are only discussing methods that benefit from privacy amplification of shuffling here. Some approaches, such as those in \cite{sun2021ldp,zhao2022privacy}, employ the shuffling step solely for breaking relations between dimensions, inherently belonging to the general CL-DP category.

Specifically, Liu et al.~\cite{liu2021flame} first consider FL in the shuffle model, proposing the FLAME algorithm. In this algorithm, each user samples the dimensions of local updates and perturbs the values in these dimensions for shuffling and aggregation. To simultaneously leverage the privacy amplification benefits of sampling and shuffling, where the privacy amplification from shuffling relies on a certain number of user samples, they further introduce the dummy padding strategy on the shuffler side, ensuring the sampling size remains constant across each dimension.  Around the different sampling strategy, Liew et al.~\cite{liew2022shuffled} proposed shuffled check-in protocol in CL-DP, which is similar to self sampling~\cite{girgis2021differentially} in the aspect of methodology but achieves $(\epsilon_l, \delta_l)$-LDP in theory and deriving a tighter privacy amplification bound with RDP. Beyond the uniform $\epsilon_l$-LDP setting for each user, Liu et al.~\cite{liu2023echo} considered heterogeneous privacy budgets that assume different $\epsilon_k$-LDP for each client, deriving a tight $(\epsilon_c, \delta_c)$-DP bound in the analyzer side based the idea of \emph{hiding among clones}~\cite{feldman2022hiding}. To improve the utility, they proposed a Clip-Laplace mechanism to bound the Laplace noise in a finite range.

\vspace{-0.2cm}

\subsubsection{Shuffle model of SL-DP}

In the shuffle model using SL-DP, the neighboring databases for $(\epsilon_c, \delta_c)$-DP analysis are defined based on samples rather than clients' data. To achieve it, a straightforward approach is to perturb each sample's updates separately, then aggregate and shuffle these samples. Chen et al. \cite{chen2023generalized} adopt this method, further exploring its privacy amplification under a heterogeneous privacy scenario and achieving a tighter privacy bound by $f$-DP (also known as GDP) compared to Liu et al.~\cite{liu2023echo}. More typically and beneficially, each user samples only one record to train and reports its corresponding local updates, as shown in Figure~\ref{figure: shuffle model}~(b).

Specifically, Girgis et al. \cite{girgis2021shuffled} introduce the CLDP-SGD algorithm for SL-DP with shuffling, which involves sampling clients first and then sampling the user's sample. Such a combination of dual sampling and shuffling enhances privacy amplification and reduces communication costs. Building on CLDP-SGD, Girgis et al. \cite{girgis2021differentially} introduce the dss-SGD algorithm. They employ Poisson sampling for client selection instead of uniformly sampling a fixed number of clients by the shuffler and demonstrate the corresponding privacy amplification effect.



\vspace{-0.2cm}

\subsection{Summary} \label{subsec:HFL-summary}
Our summary is shown in the Table~\ref{tab-Summary of HFL with DP}, which shows the object is protected and whether there are honest server in differentially private HFL. We can observe that, currently, only CL-DP assumes an honest server. Moreover, we can see that, for SL-DP, its protection object is whether a sample participates, while the protection target for CL-DP is whether a client participates. As for LDP, its protection target is the model parameters since there is no definition of neighboring datasets. For the CL-DP and LDP, a larger number of clients (cross-device) is more suitable, where the greater the number of client participants, the more the advantages of CL-DP and LDP are realized. Conversely, Cross-silo is more appropriate for SL-DP, as each client holds a larger number of samples, which better leverages the strengths of SL-DP.
Below, we explore in detail two pairs of concepts that are easily confused in DP-HFL.

\begin{table*}[h]
\vspace{-0.2cm}
\centering~
\normalsize
\caption{{Summary of DP-HFL}}
\vspace{-0.2cm}
\begin{tabular}{ccccc}
\toprule
DP Model              & Neighborhood Level & Server Assumption & Clients Setting$^{*}$& Protect Object                   \\ \hline
                                & SL                   & semi-honest      &cross-silo          & the presence of a single sample
                                \\
                                & CL                   & honest           &cross-device              & the presence of a single client
                                \\
\multirow{-3}{*}{DP}           & CL with SA           & semi-honest   &cross-device              & the presence of a single client \\ \hline
LDP                             & -                    & semi-honest  &cross-device               & the true parameters                       \\
\hline
                                & SL                   & semi-honest   &cross-silo              & the presence of a single sample \\
                            
\multirow{-2}{*}{Shuffle Modle} & CL                   & semi-honest   &cross-device              & the presence of a single client \\
\bottomrule
\end{tabular}
\begin{tablenotes}
\footnotesize
\item[*] * About Clients setting can refer~\cite{zhang2023systematic}, and what we provide here are merely appropriate client setting for the corresponding DP model, not standards.
\end{tablenotes}

\label{tab-Summary of HFL with DP}
\vspace{-0.35cm}
\end{table*}

\vspace{-0.2cm}
\subsubsection{SL-DP vs. LDP}
These two modes are often confused, and many articles categorize SL-DP as LDP because they both assume a semi-honest server. However, in fact, they are different from the definition. There is a definition of neighboring datasets in DP, and always many samples in the neighboring datasets. So, clipping is performed on each sample to limit the sensitivity brought by each sample in SL-DP. In LDP, neighboring datasets degenerate into any two different inputs, so sensitivity is often obtained by directly clipping model parameters. In terms of the number of local iterations, in SL-DP, each round of local iteration is equivalent to accessing the local dataset, requiring noise to be added to it. In LDP, the privacy loss of LDP occurs at the moment of upload, and the perturbation occurs only at the moment before upload. Moreover, many LDP articles focus more on the issue of parameter dimensions because in LDP, dimensions divide the privacy budget, leading to privacy budget explosion. Although dimensions in SL-DP also affect performance, they do not have as much impact as in LDP. Generally, LDP mechanisms can achieve the protection of SL-DP, but the reverse is not true. In special cases, such as when a client has only one piece of data, we consider SL-DP to be equivalent to LDP, but such cases are rare in the  HFL.
\vspace{-0.2cm}

\subsubsection{LDP vs. CL-DP with SA}
Both require local noise addition, but the scale of noise added locally differs. For example, using Laplace noise, in LDP, adding laplace noise $Lap(0, \epsilon/\Delta f)$, locally achieves $\epsilon$-LDP. In contrast, CL-DP only requires adding laplace noise $Lap(0, \epsilon/(K\Delta f))$, where $K$ is the number of clients, and secure aggregation ensures that the final aggregated weights satisfy $\epsilon$-DP. Although CL-DP with SA provides some level of LDP privacy protection locally, the amount of noise is too small, resulting in local privacy protection that is far less than that of $\epsilon$-LDP.

\vspace{-0.2cm}
\section{Differentially Private VFL and Differentially Private TFL} \label{sec-VerticalandTransfer}

The amount of work on DP models in VFL and TFL is far less than in HFL. In VFL, the current mainstream research focuses on secure entity alignment protocols, which involve securely matching the same entities in samples from different parties using private set intersection (PSI) algorithm~\cite{zhou2021privacy}. Then, the aligned entities can form a virtual dataset, which is used in the downstream vertical training. In TFL, current research focuses more on how to transfer knowledge from datasets across different domains and enhance model utility~\cite{fernando2013unsupervised}. There are also related articles indicating that knowledge distillation can naturally defend against some poisoning and backdoor attacks~
\cite{li2021neural}. However, as more privacy issues are exposed in VFL and TFL~\cite{pasquini2021unleashing,jagielski2024students}, research on DP models in them is becoming more and more important.

\vspace{-0.2cm}
\subsection{Differentially Private VFL} \label{subsec-VFL}

\begin{figure*}[h]
  \centering
  \begin{subfigure}{0.55\linewidth}
    \centering
		\includegraphics[width=1.0\linewidth]{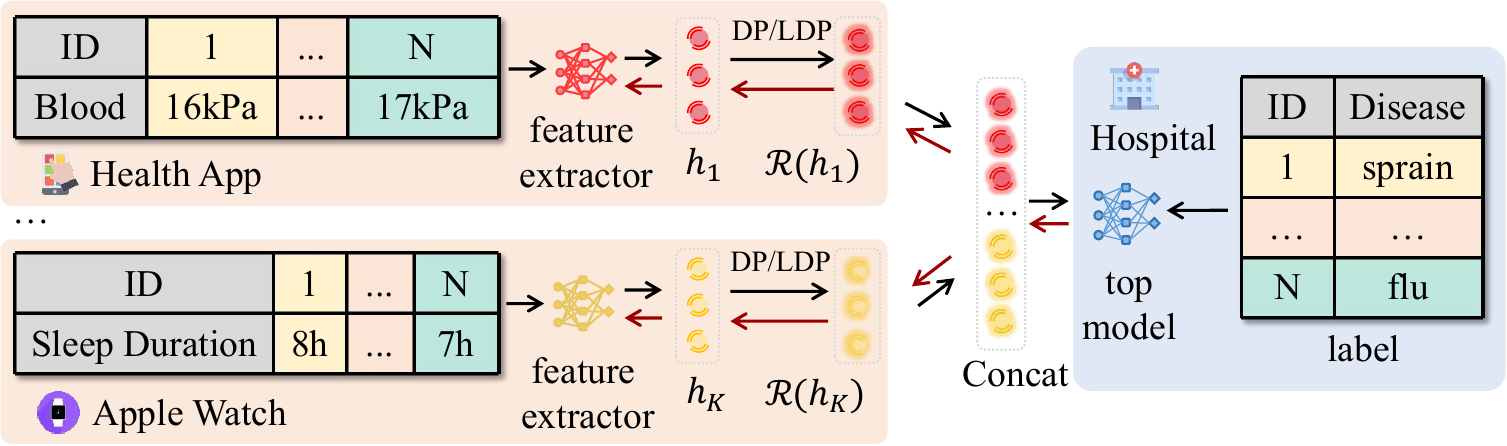}
		\caption{A example of differentially private VFL.}
		\label{fig:VFL}
  \end{subfigure}
  \hfill
  \begin{subfigure}{0.44\linewidth}
		\includegraphics[width=1.0\linewidth]{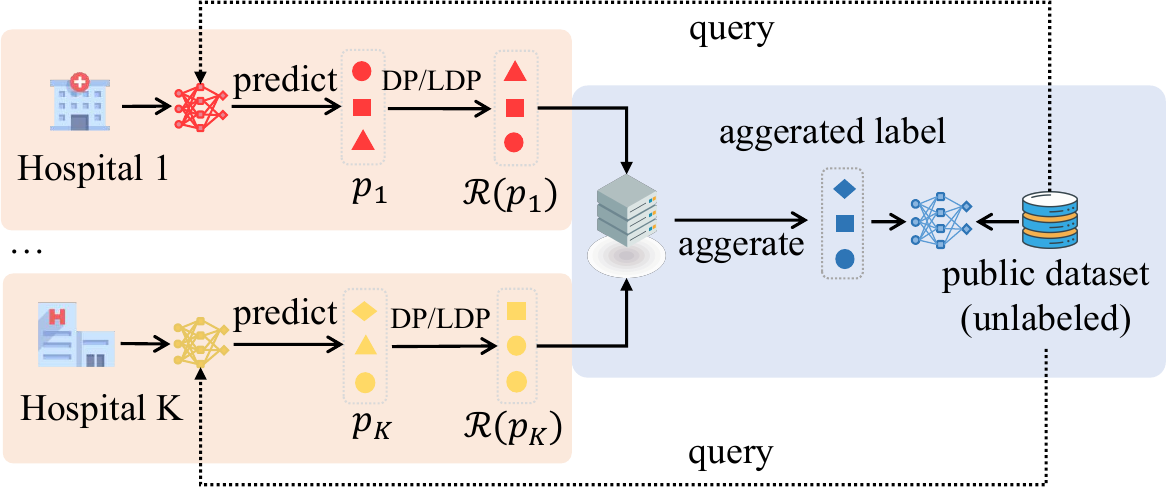}
		\caption{A example of differentially private TFL.}
		\label{fig:TFL}
  \end{subfigure}
  \caption{Differentially private VFL and TFL.}
  \label{figure: VFL and TFL}
  \vspace{-0.4cm}

\end{figure*}

VFL is suitable for joint training when different clients hold the same samples with different features, increasing the feature dimensions of data during training \cite{yang2019federated}. As shown in Figure \ref{fig:VFL}, there are two apps collecting user health data in a region: one is the Health App, and the other is the Apple Watch. It is highly likely that the information of local residents is registered in both of these apps, while the disease information is only held by the Hospital. However, the user features may have no commonalities, as the Health App records users’ blood pressure, while the Apple Watch records users’ sleep duration. In the case of aligned data samples, all parties can engage in joint training. However, this joint training method poses a serious privacy threat, as both features and labels can potentially be inferred. A common protective method is to add perturbation $\mathcal{R}(h_k)$ to the features $h_k$ extracted by the feature extractor for DP protection.

Clients transmit prediction results to the server, posing privacy risks of leaking features \cite{pasquini2021unleashing}. Therefore, some existing work has introduced DP into the model training of VFL to protect features. Chen et al. \cite{chen2020vafl} introduced the concept of asynchrony based on VFL. Although there are multiple clients, only one client and server are active at the same time. They introduced noise by ``increasing random neuro'' and controlled the variance of Gaussian random neurons to satisfy Gaussian differential privacy (GDP). Oh et al. \cite{oh2022differentially} replaced traditional neural network models with a vision transformer (ViT) and used split learning (SL) to bypass the large size issue of ViT by shattering data communication at the slicing layer. They also proposed DP-CutMixSL, introducing a Gaussian mechanism to the data smashed by cutout uploaded to the server, making the algorithm satisfy DP. Mao et al. \cite{mao2022secure} proposed a new activation function called ReLU, which transforms private shuffled data and partial losses into random responses in forward and backward propagation to prevent attribute inference and data reconstruction in SplitNN. Based on this, they introduced differential privacy and added Laplace noise to the forward propagation results in a random response (RR) manner, achieving a fine-grained privacy budget allocation scheme for SplitNN. Tian et al. \cite{tian2023sf} introduced "bucket" into the scenario of combining VFL with gradient boosting decision tree (GBDT), allowing the server to only know which bucket a sample belongs to, without knowing the order of the sample in the buckets. By moving samples between buckets with a certain probability and adding RR perturbation to each bucket, the samples are protected by DP. Li et al. \cite{li2022opboost} proposed condensed-LDP (CLDP) in the context of VFL combined with tree boosting and designed three order-preserving desensitization algorithms for feature information to achieve privacy protection. Liao et al. \cite{liao2025privacy} proposed MFVFL, a tensor decomposition-based vertical federated learning framework that resolves feature missingness through client learnable imputation and server low-rank decomposition, while its privacy-preserving variant MFVFL-DP employs tensor robust PCA to mitigate differential privacy noise impact.

On the other hand, relevant research indicates that privacy information of labels can be inferred based on the gradients returned by the server~\cite{fu2022label,li2021label}. Therefore, some existing work has introduced DP into VFL to protect labels of server. Yang et al.~\cite{yang2022differentially}  proposed a label-protected VFL, where Laplace noise is added to the gradients obtained by the server through reverse derivation to protect label privacy information, as the server needs access to label information when computing gradient values. On the other hand, Takahashi et al.~\cite{takahashi2023eliminating} applied Label Differential Privacy to VFL. When calculating the loss value, they directly perturb the true labels using General Random Response (GRR) to protect label information. Their method successfully mitigated label leakage from the instance space, effectively countering ID2Graph attacks based on tree models.

In addition, related research simultaneously protects both features and labels in VFL. Wang et al. \cite{wang2020hybrid} not only add Gaussian noise to the results of forward propagation obtained by the feature holder (client), but also add Gaussian noise to the gradient results obtained by the label holder (server) during backward derivation, thus ensuring differential privacy protection on both sides. Wu et al. \cite{wu2020privacy} used MPC to construct a trusted third-party environment for centralized training, enabling the application of differential privacy in a dense environment, ensuring that the final output model satisfies differential privacy protection.

\vspace{-0.2cm}

\subsection{Differentially Private TFL} \label{subsec-TFL}
TFL is applicable to data holders who do not possess or only possess a small number of identical samples and features, enabling them to perform joint training. As shown in Figure \ref{fig:TFL}, there are $K$ hospitals located in different regions, such as Hospital $1$ in Japan, ..., and Hospital $K$ in the United States. Due to geographic limitations, the user populations of multiple medical entities have little overlap, and the data features of these entity datasets hardly overlap. In order to transfer the knowledge of the local models of these hospitals to unlabeled public datasets, Knowledge Distillation (KD) \cite{hinton2015distilling} is commonly used, which achieves knowledge transfer by querying and aggregating the output predictions of each local model. However, relevant research has shown that attackers can attack the original data information based on the output predictions of the local models \cite{zhang2020secret}. Therefore, numerous studies aim to incorporate differential privacy techniques into TFL in order to enhance data privacy protection.

Papernot et al. \cite{papernot2017semi} first introduced DP into TFL and proposed the Private Aggregation of Teacher Ensembles (PATE) algorithm. It divides the private dataset into multiple disjoint data batches and trains a ``teache'' model for each batch. Then, each teacher submits its predictions for the public unlabeled dataset (student) to an aggregator. The aggregator collects and perturbs the predictions of all teachers to ensure that the aggregated labels maintain differential privacy, and sends the perturbed predictions to the students for training. The following year, Papernot et al. \cite{papernot2018scalable} optimized the PATE algorithm by using Gaussian mechanism for perturbation and obtaining tighter privacy bounds using RDP, making PATE applicable on a larger scale. Tian et al. \cite{tian2022seqpate} applied PATE to text generation models. To reduce the noise required by DP for large output spaces in vocabulary size, they dynamically filter out unimportant words to reduce the output space. They also proposed an effective knowledge distillation strategy to reduce the number of queries. Abourayya et al.~\cite{abourayya2025little} apply consensus methods like majority voting to generate pseudo-labels from hard labels, replacing traditional soft label sharing. Protecting these hard labels via DP enhances privacy protection.

However, traditional PATE methods only have one client's private dataset. Pan et al. \cite{pan2021fl} extended it to multiple clients and treated each client as a teacher model, achieving client-level DP with a trusted server. To address the problem of highly inconsistent predictions among teachers due to data heterogeneity among multiple clients, Dodwadmath et al. \cite{dodwadmath2022preserving} improved PATE by using an auxiliary global model that incorporates teacher averaging and using update correction to reduce the variance of teacher updates. To eliminate the reliance on a trusted third party, Qi et al. \cite{qi2023differentially} converted the predictions outputted by each client teacher into one-hot encoding for prediction and perturbed each one-hot value using Random Response (RR) to satisfy LDP.

In addition, researchers have started to study privacy protection scenarios where there is a small overlap of user data among the participants. Wan et al. \cite{wan2023fedpdd} proposed the privacy-preserving dual distillation framework called FedPDD. Clients use locally trained models to make predictions on a small amount of overlapping user data and submit the prediction results to the server. The server aggregates the local predictions of the participants to obtain ensemble teacher knowledge and distributes it to the parties for subsequent training. Differential privacy protection is applied during the aggregation of local predictions on the server. Sun et al. \cite{sun2020federated} use sampling with replacement or without replacement of participant data samples to achieve noiseless differential privacy. Hoech et al. \cite{hoech2022fedauxfdp} proposed the FedAUXfdp algorithm, which adds Gaussian noise to the regularized multinomial logistic regression vector during local client training of classification models, upgrades the feature extractor to a frozen feature extractor, and solves the problem of heterogeneous client data distribution in one-shot scenarios while satisfying differential privacy.

\vspace{-0.2cm}
\section{Applications} \label{sec-application}
For the most part, the techniques discussed in previous sections have been demonstrated on relatively standard tasks (e.g. image classification on benchmark datasets, as summarized in Table~\ref{tab: An overview study of DP-FL}). In this section, we broaden the scope by reviewing how differentially private federated learning is being applied across diverse data types and real-world scenarios. We also provide deeper analysis of the unique challenges and solutions in these application domains.

\vspace{-0.2cm}
\subsection{Applications across Data Modalities.}

\subsubsection{Graph Data}
Recent advancements in FL applied to graph data focus on developing methodologies that safeguard privacy without sacrificing the utility of graph neural networks (GNNs). The core operation of GNNs is message passing, which makes the learning of a node's embedding depends not only on the node itself but also on its neighboring nodes. Accurately measuring this dependency level is crucial for determining sensitivity, which determines the noise scale introduced by the DP mechanism~\cite{daigavane2022node}. So, the main challenge in applying graph data is controlling the depth and breadth of propagation to limit sensitivity while utilizing the transitivity within graph data, thereby minimizing the impact of DP-introduced noise on model utility.
Currently, methods to address the dependency issue can be divided into two categories. The first category avoids direct measurement of the dependency level by adding noise outside the GNN training phase. Wu et al. proposed LinkTeller, a link-based attack for the vertical federated learning scenario, and implemented link privacy protection by adding Laplacian noise to the graph's adjacency matrix~\cite{wu2022linkteller}. Another approach proposed in~\cite{lin2022towards} perturbs decentralized graph data by protecting edge and node features of each user's adjacency list through randomized response. Under the same setting, Lin et al. suggested perturbing locally trained node embeddings~\cite{wu2021fedgnn}. However, these perturbation mechanism results in significant usability loss of the model since the design is not tailored to GNNs. Sajadmanesh et al. proposed extracting the aggregation function from the GNN, performing manual message passing, and adding noise to the aggregated message~\cite{sajadmanesh2023gap}. This method avoids complex dependency measurement but renders the aggregation function non-learnable, limiting its application to other network architectures. The second category measures the dependency level of nodes and applies gradient perturbation for private GNN training. Daigavane et al. proposed limiting the in-degree of nodes through sampling, measuring the upper bound of the dependency level, and implementing node differential privacy using the DP-SGD paradigm~\cite{dai2021differentially}. According to the theoretical results, the sensitivity remains high, especially as the depth of GNN layers increases. proposed a federated graph analysis framework, FEAT, that supports arbitrary downstream graph statistics while protecting individual privacy. Liu et al.~\cite{liu2024federated}
proposed a federated graph analysis framework, that supports arbitrary downstream graph statistics while satisfying DP guarantee.

\vspace{-0.2cm}
\subsubsection{Bipartite Interaction}
In a bipartite matrix, rows and columns represent different entities, and the elements describe the relationships between these entities. This type of data is a special case of graph data, but it needs specific discussion due to its relevance to the recommendation systems. In recommendation systems, the rows and columns of a bipartite matrix represent users and items, respectively. The observed data can be categorized into two types: explicit feedback, such as 1-5-star ratings reflecting the intensity of user preferences, and implicit feedback, such as retweets and clicks, which indicate interactions without explicitly reflecting preferences~\cite{koren2021advances}.

Studies such as~\cite{friedman2016differential} primarily focus on scenarios involving explicit feedback. They propose a DP protection framework for explicit data in bipartite matrices using matrix factorization models as the building block. This framework applies input, process, and output perturbation mechanisms, with results indicating that process perturbation mechanisms based on stochastic gradient descent and alternating least squares achieve the best outcomes. Other studies, such as~\cite{minto2021stronger} and~\cite{ammad2019federated}, explore federated collaborative filtering with implicit feedback, employing user-level differential privacy and highlighting that greater noise must be added to meet unbounded DP requirements with implicit feedback. Bipartite matrices, used in recommendation systems with many items and a sparse structure, pose communication challenges in federated learning due to high overhead. Shin et al.~\cite{shin2018privacy} proposed solution involves clients uploading only a single element of the gradient matrix based on a randomized selection mechanism rather than the entire gradient matrix. This allows the server to aggregate unbiased estimates of the global gradient matrix, ensuring the algorithm satisfies LDP while maintaining the usability of the gradient matrix.

\vspace{-0.2cm}
\subsubsection{Time Series Data}
In time series data scenarios, unlike the single release of data from a static database, adversaries can observe multiple differentially private outputs. 
To publish time series data with DP in FL settings, a widely accepted approach is to use a hierarchical structure. The idea is to partition the time series into multiple granularities and then add noise to the time series to satisfy DP~\cite{chan2011private,dwork2010differential}. The size of the DP noise is generally determined by the upper bound of the data. However, time series data is often concentrated below a value much smaller than the upper bound. This issue is typically addressed using truncation-based methods~\cite{perrier2018private}. Wang et al. optimized the searching for the truncating threshold. Their contributions include an EM-based algorithm to find the threshold and an online consistency algorithm~\cite{wang2021continuous}. 

On the other hand, handling time series data often requires using recurrent neural networks (RNNs) and their variants for modeling. Typically, when there are not many participants, using some RNN variants is feasible under DP, e.g., gated recurrent unit neural networks~\cite{xu2022edge}. However, for tasks like traffic flow forecasting, it is difficult to converge due to many participants and expensive communication overhead. To address this, Liu et al. designed a joint-announcement protocol in the aggregation mechanism to randomly select a certain proportion of organizations from many participants in the $i$-th round of training~\cite{liu2020privacy}. For time series data related to location information, such as trajectory prediction, better spatiotemporal correlation of the data leads to better performance. Thus, leveraging spatiotemporal correlation is crucial. Liu et al. pre-clustered clients based on this principle. It integrates the global model of each cluster center using an ensemble learning scheme, thereby achieving the best accuracy~\cite{liu2020privacy}. The clustering decision is determined by using the latitude and longitude information.


\vspace{-0.2cm}
\subsection{Applications in Real-World Scenarios}

\subsubsection{Health Medical}

Recent studies have focused on enhancing privacy in the application of FL within the healthcare domain, particularly addressing the concerns around the confidentiality of patient data when leveraging the Internet-of-Medical-Things (IoMT) and distributed healthcare datasets. The research highlighted in~\cite{wu2021incentivizing} introduces the concept of adding artificial noise to IoMT device datasets for user privacy. Similarly,~\cite{malekzadeh2021dopamine} adopts a differentially private stochastic gradient descent approach, combined with secure aggregation through homomorphic encryption, to work on distributed healthcare data, demonstrating its efficacy on a dataset of diabetic retinopathy. Meanwhile,~\cite{choudhury2019differential} explores the use of differential privacy in FL to model Electronic Health Records (EHR) across various hospitals, focusing on tasks such as predicting adverse drug reactions and mortality rates, with a significant dataset containing sensitive information like diagnosis and admission records. Additionally, Zhou et al.~\cite{zhou2024ppml} extended differentially private federated learning to genomics data analysis tasks (such as cancer classification with bulk RNA-seq), thereby protecting patients' private information.


\vspace{-0.2cm}

\subsubsection{Internet of Things}
FL is explored extensively in the IoT recently. Lu et al.~\cite{lu2019differentially} leverage FL to protect
the privacy of mobile edge computing and they propose a random distributed update scheme to get rid of the security threats led by a centralized curator, while Pan et al.~\cite{pan2021joint}
apply DP to energy harvesting with collaborative and intelligent protection cross the energy side and information side. 
He et al.~\cite{he2023clustered} propose a LDP scheme to train clustered FL models on heterogeneous IoT data by utilizing dimension reduction methods. Tao et al.~\cite{tao2024private} combined random sparsification with Gaussian perturbation to achieve DP privacy guarantees and bandwidth-adaptive communication.


\vspace{-0.2cm}

\subsubsection{Finance and Risk Modeling}
Differentially private federated learning has gained traction in finance, powering collaborative credit scoring, fraud detection, and risk analytics without exposing sensitive customer data. For example, the FDML framework introduced by Hu et al.~\cite{hu2019fdml} enabled banks to jointly train credit risk models on feature-partitioned data. Similarly, the Helen system developed by Zheng et al.~\cite{zheng2019helen} demonstrated secure multi-party fraud detection via cryptographic protocols. Nevertheless, privacy threats remain significant. Melis et al.~\cite{melis2019exploiting} showed that federated learning model updates can leak user behavior patterns. Fu et al.~\cite{fu2022label} further revealed the risk of cross-silo label inference attacks. These findings underscore the core challenges of DP-FL: the injected noise exacerbates the imbalance caused by non-IID financial data. To address these issues, emerging solutions, such as the one proposed by Chowdhury et al.~\cite{roy2020crypt}, combine differential privacy with secure computation to enhance privacy in multi-institutional analytics.

\vspace{-0.2cm}


\section{Open Challenges and Future Direction} \label{sec-future direction}
In addition to achieving a better trade-off between privacy protection and model utility in differentially private federated learning, we present some challenges in current studies and provide promising directions for future research.

\vspace{-0.2cm}
\subsection{Convergence Analysis of Differentially Private HFL}
In HFL, A tight upper bound on convergence not only provides theoretical assurance of rapid convergence, but also enables an analysis of the impact of various hyperparameters on the speed of convergence, which can guide parameter tuning or the proposal of new optimization algorithms~\cite{haddadpour2019convergence,li2024convergence}. However, the convergence analysis of differentially private HFL, compared to HFL, needs to consider the effects of operations such as clipping and adding noise, making it more challenging to obtain a relatively tight upper bound \cite{zhang2022understanding}. Wei et al. \cite{wei2020federated} introduced the NbAFL algorithm and provided its convergence analysis, while Ling et al. \cite{ling2024ali} offered convergence analysis for DPSGD across multiple iterations, but their assumptions about the objective function were strong. Three studies \cite{zhang2022understanding,cheng2022differentially,shi2023make} conducted convergence analysis of their respective CL-DP algorithms under non-convex conditions, addressing the issue of overly strong assumptions on the objective function but still making strong assumptions on gradients, such as Bounded Gradient and Bounded Variance.

So, conducting convergence analysis without strong assumptions on the loss function and gradients in differentially private HFL is a challenging. Furthermore, current analyses of the proposed differentially private HFL algorithms’ convergence upper bound remain at $\mathcal{O}(\frac{1}{\epsilon})$~\cite{li2022soteriafl}, finding tighter convergence bounds is a future direction.
\vspace{-0.2cm}

\subsection{User-Level DP in HFL}
We have discussed the sample-level and client-level of DP as above, but in many scenarios, there are many users in a client and individual user may contribute multiple samples or records. For example, in a hospital, a patient may have multiple medical records. So, For each hospital, the goal of user-level DP is to hide the presence of a single user, or to be more specific, to bound the influence of any single user on the learning outcome distribution (i.e. the distribution of the model parameters). So, in a user-level DP scenario, we need to capture the collection of all data points associated with a user. We can consider a dataset with $m$ users and $n$ samples, where $m \leq n$.  Therefore, individual users can have multiple data samples. Therefore, based on the differential neighboring datasets held by data owners, we can get the notion of UL-DP in
HFL as follows.

\vspace{-0.2cm}
\begin{definition} \label{def-sl}
    ({\bf User-level DP (UL-DP)}). Under UL-DP, two datasets $D$ and $D^{\prime}$ are neighbouring if they differ in all samples (records) from a single user. (either through addition or through removal).
\end{definition}
\vspace{-0.2cm}

In UL-DP, allowing users to contribute a large number of data samples, even if most contribute only a few, may eventually introduce excessive noise to protect against minority outliers. On the other hand, restricting users to make only small contributions can maintain a lower level of noise but may discard a large amount of surplus data, introducing bias. Some related research has investigated user-level considerations in tasks such as mean estimation, linear regression and experience risk minimization~\cite{epasto2020smoothly,liu2020learning,levy2021learning,kato2024uldp}. Although \cite{kato2024uldp} represents the pioneering work on UP-DP in FL, this field is still underexplored.
\vspace{-0.2cm}

\subsection{Privacy Auditing for various DP Models and Neighborhood Levels in FL}
Above, we have discussed various DP models and neighborhood levels in federated learning. In particular, as shown in Table~\ref{tab-Summary of HFL with DP}, we can observe that the objects directly protected by differential privacy federated learning vary based on different DP models and neighborhood levels. Currently, many articles have been devoted to algorithm design based on different DP models and neighborhood levels, resulting in different perturbation algorithms. However, there has not yet been an article that provides a comprehensive comparison among them. Although privacy protection can be characterized by privacy budget $\epsilon$, these definitions may have different privacy protection effects in practical scenarios due to their diverse objects of protection. 

Therefore, conducting practical attack methods to quantify their privacy effects is a important direction of future work. For instance, utilizing membership inference attacks~\cite{salem2019ml}, attribute inference attacks~\cite{melis2019exploiting}, and data reconstruction attacks~\cite{zhao2020idlg} to measure their impacts. Or using DP auditing technology~\cite{jagielski2020auditing,nasr2023tight} to approach the worst case guarantee under various DP models and levels, for example, LDP and SL-DP. Although Naseri et al.~\cite{naseri2020local} have conducted related research, comparing SL-DP with CL-DP based on a trusted server, we believe their comparison is not comprehensive enough and that these two definitions cannot be directly compared due to their different security assumptions and objects of protection. Secondly, DP guarantee based on implausible worst-case assumptions, this makes it difficult for the privacy budget $\epsilon$ to intuitively correspond to real-world privacy leakage scenarios. So, it is valuable to discuss the gap between DP guarantee under various DP models and neighborhood levels and ML privacy attacks~\cite{salem2023sok}.


\vspace{-0.2cm}
\subsection{Differentially Private Federated Learning for Fine-tuning Large Language Models (LLMs)}
There is significant potential in using federated learning to fine-tune Large Language Models (LLMs) with differential privacy on local data to preserve privacy~\cite{pan2025selective}. For example, Wang et al. ~\cite{wang2023can} leverages public pre-trained LLMs and introduces a distribution matching algorithm to improve sample efficiency in public training, showcasing a strong privacy-utility trade-off without relying on pre-trained classifiers. Meanwhile, Xu et al.~\cite{xu2023federated} implements DP-FTRL in Google Keyboard (Gboard), ensuring formal differential privacy guarantees without uniform client device sampling. However, it faces considerable challenges. One major challenge is the substantial communication overhead that arises when applying federated learning to LLM fine-tuning. This is due to the iterative aggregation of model updates from decentralized sources. The frequent transmission of these updates for a complex model like an LLM is extremely costly. While some strategies exist to mitigate these communication challenges, a notable trade-off persists: these methods alleviate communication bottlenecks but cannot provide concrete guarantees for protecting training data privacy.

To address this issue, one can design a communication-efficient federated learning approach with differential privacy guarantees, such as DP-LoRA~\cite{liu2024differentially}. It leverages the inherent low-rank properties of LLMs by integrating a small set of new weights into the model. By focusing fine-tuning on only these parameters, it drastically reduces communication costs, as only a limited number of updated weights need to be transmitted during distributed training~\cite{liu2025differentially}. Besides these, 
Differentially private zeroth-order optimization (DP-ZO) has shown promise in fine-tuning LLMs while protecting privacy~\cite{bao2025unlocking}, DP-ZO replaces the exact first-order gradient (derived via backpropagation) with a random zero-order approximation based on querying the model's loss, which significantly accelerates model training. How to apply DP-ZO to federated fine-tuning LLMs is also a significant but challenging problem.

Another important challenge is the trade-off between model utility and privacy budget. DPSGD protects privacy by adding noise to model updates. However, the immense number of parameters in the embedding layers of language models requires a substantial amount of noise. This severely impacts the model's learning ability, a phenomenon known as the curse of dimensionality~\cite{de2022unlocking}. A feasible solution is to leverage public data for pre-training and knowledge distillation. One approach is to conduct a large-scale, non-private pre-training on public data, and then fine-tune on the private data. Alternatively, knowledge distillation can be used to distill the knowledge learned from private data into another model, which bypasses the challenge of adding a large amount of noise directly to sensitive, high-dimensional data.

\vspace{-0.2cm}
\subsection{Cross-domain differentially private TFL}
Current works of differentially private TFL mostly focus on intra-domain scenarios, where the source and target datasets share similar features. However, in real-world scenarios, the source and target datasets often exhibit different features (cross-domain)~\cite{fernando2013unsupervised,pan2010domain}. Unsupervised Multi-source Domain Adaptation (UMDA) addresses this by transferring transferable features from multiple source domains to an unlabeled target domain~\cite{zhang2015multi,chang2019domain}. Due to the unavailability of direct access to sensitive data, all data and computations on the source domain must remain decentralized. Additionally, since uploaded parameters may leak sensitive information, we need to protect the parameters uploaded from the source domain with differential privacy. Therefore, striking a balance between the privacy and utility of cross-domain differentially private federated learning poses a key challenge currently.

\subsection{Multimodal Federated Learning with Differential Privacy}
Federal learning has been widely applied to multimodal settings (e.g., multi-sensor or vision-language tasks)~\cite{che2023multimodal}. However, there is a lack of work combining multimodal federated learning with DP to protect complex data combinations. The challenges faced include: complementary cross-modal information can lead to inference attacks, and varying modality dimensions complicate the noise calibration for differential privacy. These issues can severely degrade model utility; for example, DPSGD often fails to converge on high-dimensional multimodal models. Therefore, to mitigate the performance loss caused by differential privacy, effective differentially private federated learning for multimodal data must balance high-dimensional sensitivity and data heterogeneity while preventing cross-modal information leakage.

\section{Conclusion} \label{sec-conclusion}
In this study, we explored and systematized the differentially private federated learning. 
We categorized DP models in FL into three major classes, namely DP, LDP, and shuffle model, based on the definitions and guarantees of differential privacy. Further, within DP and shuffle model, we differentiated between SL-DP and CL-DP based on the definition of neighboring datasets. Subsequently, we showed the applications of differentially private federated learning in differential data types and real-world scenarios. Based on these discussions, we provided 6 promising directions for future research. We aim to provide practitioners with an overview of the current technical and application of differential privacy in federated learning, stimulating both foundational and applied research in the future.



\appendix
\vspace{-0.2cm}
\section{Loss Composition Mechanisms in DP}\label{sec:Loss Composition Mechanisms in DP}

There are often multiple rounds of iteration required to obtain the model in differentilly private FL, where each iteration can be considered as a data access. By designing a differentially private algorithm $\mathcal{A}_{i,i \in [k]}$ that satisfies $(\epsilon,\delta)$-DP for the $k$ iteration, we can leverage the sequential composition theorem to compute the privacy budget $(k\epsilon,k\delta)$-DP for the entire process. But, this is not tight and
we can use the advanced composition that could be improved to $(O(\sqrt{k\epsilon}),O(k\delta))$-DP~\cite{dwork2010boosting}. 

\textbf{More privacy loss composition mechanisms for DP.}
However, the number of iterations required for model convergence in differentilly private FL is often enormous. The advanced composition theorem no longer guarantees a more compact estimate of privacy loss, prompting the emergence of more compact measures for combining privacy loss, among which commonly used ones include Zero-Concentrated Differential Privacy (zCDP)~\cite{bun2016concentrated}, Moments Accountant (MA)~\cite{abadi2016deep}, Rényi Differential Privacy (RDP)~\cite{mironov2017renyi} and Gaussian Differential Privacy (GDP)~\cite{dong2021gaussian}. All these privacy composition measures mechanisms achieve a tighter analysis of cumulative privacy loss by leveraging the fact that the privacy loss random variable is strictly centered around the expected loss. It is important to note that these mechanisms are just different methods for analyzing privacy composition, and their purpose is to achieve a tighter analysis of the guaranteed privacy. And all these mechanisms can be converted to $(\epsilon,\delta)$-DP. This means that for a fixed privacy budget $(\epsilon,\delta)$, a relatively loose definition can be satisfied by adding relatively less noise, thus achieving less privacy for the same $(\epsilon,\delta)$ level. In practice, we often utilize these composition mechanisms to combine privacy losses and then convert them to $(\epsilon,\delta)$-DP for comparison.



The Moments Accountant (MA) technique~\cite{abadi2016deep} used cumulative generating function (CGF) to characterize the privacy bounds of two privacy random distributions. The definition of CGF is as follows:
\vspace{-0.2cm}
\begin{definition}({\bf CGF \cite{abadi2016deep}}) Given two probability distributions $P$ and $Q$, the CGF of order $\alpha > 1$ is: 
\begin{align}
	\begin{split}
		\begin{aligned}
G_{\alpha}(P || Q) =  & \operatorname{log}\mathbb{E}_{x\sim P(x)}\bigg[e^{\alpha \operatorname{log}\frac{P(x)}{Q(x)}}\bigg]
        = \operatorname{log} \mathbb{E}_{x \sim Q(x)}\left[\left(\frac{P(x)}{Q(x)}\right)^{\alpha+1}\right]
		\end{aligned}
	\end{split}
\end{align}
where $\mathbb{E}_{x \sim Q(x)}$ denotes the excepted value of $x$ for the distribution $Q$, $P(x)$, and $Q(x)$ denotes the density of $P$ or $Q$ at $x$ respectively.
\end{definition}
\vspace{-0.2cm}

Based on CGF, the definition of privacy bounds for MA tracking is as follows:
\vspace{-0.2cm}

\begin{definition}({\bf MA \cite{abadi2016deep}}).
For any neighboring datasets $D, D^\prime \in \mathcal{X}^n$ and all $\alpha\in(1,\infty)$, a randomized mechanism $\mathcal{A}: \mathcal{X}^n \rightarrow \mathbb{R}^{d}$ satisfies $(\epsilon,\delta)$-DP if
\begin{align}
    G_{\alpha}(\mathcal{A}(D) || \mathcal{A}(D^\prime)) \leq \epsilon, \,\, \text{where} \,\,\delta=\min_{\alpha}e^{G_{\alpha}(\mathcal{A}(D) || \mathcal{A}(D^\prime))-\alpha\epsilon}.
\end{align}

\end{definition}
\vspace{-0.2cm}

Rényi divergence is another metric that measuring distinguishability of two random distributions, defined as follows:
\vspace{-0.4cm}
\begin{definition}({\bf Rényi Divergence \cite{van2014renyi}}). Given two probability distributions $P$ and $Q$, the Rényi divergence of order $\alpha > 1$ is: 
\begin{align}
D_{\alpha}(P \| Q)=\frac{1}{\alpha-1} \ln \mathbb{E}_{x \sim Q(x)}\left[\left(\frac{P(x)}{Q(x)}\right)^{\alpha}\right],
\end{align}
where $\mathbb{E}_{x \sim Q(x)}$ denotes the excepted value of $x$ for the distribution $Q$, $P(x)$, and $Q(x)$ denotes the density of $P$ or $Q$ at $x$ respectively.
\end{definition}
\vspace{-0.2cm}

Based on Rényi divergence, zCDP can be obtained as follows:

\vspace{-0.2cm}
\begin{definition}({\bf zCDP \cite{bun2016concentrated}})
For any neighboring datasets $D, D^\prime \in \mathcal{X}^n$ and all $\alpha\in(1,\infty)$, a randomized mechanism $\mathcal{A}: \mathcal{X}^n \rightarrow \mathbb{R}^{d}$ satisfies $R$-zCDP if
\begin{align}
    D_{\alpha}(\mathcal{A}(D) || \mathcal{A}(D^\prime)) \leq R\alpha.
\end{align}
\end{definition}
\vspace{-0.2cm}
And the following Lemma~\ref{lem:conversion zCDP to DP} defines the standard form for converting $(\alpha, R)$-zCDP to ($\epsilon$, $\delta$)-DP.
\vspace{-0.2cm}
\begin{lemma}\label{lem:conversion zCDP to DP}
({\bf Conversion from zCDP to DP~\cite{Bun_Steinke_2016}}). if a randomized mechanism $A : D \rightarrow \mathbb{R}$  satisfies $(\alpha,R)$-zCDP ,then it satisfies$(R+2\sqrt{R\log(1/\delta)}, \delta)$-DP for any $0<\delta<1$.
\end{lemma}
\vspace{-0.2cm}

Rényi differential privacy (RDP) is also based on the definition of Rényi divergence as follows:
\vspace{-0.2cm}
\begin{definition}({\bf RDP \cite{mironov2017renyi}}) For any neighboring datasets $D, D^\prime \in \mathcal{X}^n$, a randomized mechanism $\mathcal{A}: \mathcal{X}^n \rightarrow \mathbb{R}^{d}$ satisfies $(\alpha, R)$-RDP if
\begin{align}
    D_{\alpha}(\mathcal{A}(D) || \mathcal{A}(D^\prime)) \leq R.
\end{align}
\end{definition}
\vspace{-0.2cm}
And the following Lemma~\ref{lem:conversion RDP to DP} defines the standard form for converting $(\alpha, R)$-RDP to ($\epsilon$,  $\delta$)-DP.
\vspace{-0.2cm}
\begin{lemma}\label{lem:conversion RDP to DP}
({\bf Conversion from RDP to DP~\cite{balle2020hypothesis}}). if a randomized mechanism $A : D \rightarrow \mathbb{R}$  satisfies $(\alpha,R)$-RDP ,then it satisfies$(R+\ln ((\alpha-1) / \alpha)-(\ln \delta+ \ln \alpha) /(\alpha-1), \delta)$-DP for any $0<\delta<1$.
\end{lemma}
\vspace{-0.2cm}

Dong et al.~\cite{dong2021gaussian} used hypothesis testing to quantify the distinguishability between $\mathcal{A}(D)$ and $\mathcal{A}(D^\prime)$.
They considerd a hypothesis problem $H_0:P\text{ v.s. }H_1:Q$ and a rejection rule $\phi\in[0,{1}]$. They defined the type I error as $\alpha_{\phi}=\mathbb{E}_{P}[\phi]$, which is the probability that rejecting the null hypothesis $H_0$ by mistake. And the type II error $\beta_{\phi}=1-\mathbb{E}_{Q}[\phi]$ is the probability that accepting the alternative $H_1$ wrongly. And the trade-off function aims to minimal type II error at level $\alpha$ of the type I error as follows.
\vspace{-0.2cm}
\begin{definition}({\bf Trade-off function \cite{dong2021gaussian}}). Given two probability distributions $P$ and $Q$, the trade-off function of them is:
\begin{align}
    T(P,Q)(\alpha)=\inf_\phi\{\beta_\phi:\alpha_\phi\leq\alpha\}.
\end{align}
\end{definition}
\vspace{-0.2cm}

Let $f$ be a trade-off function. Algorithm $\mathcal{A}$ is the $f$-differentially private if $T(\mathcal{A}(D),\mathcal{A}(D^{\prime}))\ge f$ for two neighboring datasets $D$ and $D^{\prime}$~\cite{dong2021gaussian}. When the trade-off function is defined between two Gaussian distributions, can derive a subfamily of $f$-differential privacy guarantees called GDP as follows.

\vspace{-0.2cm}
\begin{definition}({\bf GDP \cite{dong2021gaussian}}) Let $\Phi$ denote the cumulative distribution function of the standard normal distribution. For any neighboring datasets $D, D^\prime \in \mathcal{X}^n$ and $\mu \ge 0$, a randomized mechanism $\mathcal{A}: \mathcal{X}^n \rightarrow \mathbb{R}^{d}$ satisfies $\mu$-GDP if
\begin{align}
    T(\mathcal{A}(D),\mathcal{A}(D^\prime))\geq G_{\mu},
\end{align}
where $G_{\mu}:=T(\mathcal{N}(0,1),\mathcal{N}(\mu,1))\equiv\Phi(\Phi^{-1}(1-\alpha)-\mu)$.
\end{definition}
\vspace{-0.2cm}

And the following Lemma~\ref{lem:conversion GDP to DP} defines the standard form for converting $\mu$-GDP to ($\epsilon$,  $\delta$)-DP.
\vspace{-0.2cm}
\begin{lemma}\label{lem:conversion GDP to DP}
({\bf Conversion from GDP to DP~\cite{dong2021gaussian}}). if a randomized mechanism $A : D \rightarrow \mathbb{R}$ satisfies $\mu$-GDP, then it satisfies $(\epsilon,\Phi(-\frac{\varepsilon}{\mu}+\frac{\mu}{2})-\mathrm{e}^{\varepsilon}\Phi(-\frac{\varepsilon}{\mu}-\frac{\mu}{2}))$-DP for any $\epsilon > 0$.
\end{lemma}
\vspace{-0.2cm}

Another popular property is subsampling, which achieves privacy amplification by running the differentially private algorithm on a subset of the privacy samples instead of all of them~\cite{li2012sampling}. And the privacy amplification guarantee is different for different subsampling methods~\cite{imola2021privacy,zhu2019poission}.

 \vspace{-0.2cm}

\section{Fundamental Mechanisms for Differentially Private FL} \label{sec:Fundamental Mechanisms for differentially private FL}

In this section, we introduce the fundamental perturbation mechanisms used in differentially private FL. As depicted in Table~\ref{table:Fundamental perturbation mechanisms for DP-FL}, these mechanisms are categorized along two dimensions: DP or LDP, and continuous or discrete data types. Our classification focuses on their primary application scenarios in FL. Notably, mechanisms such as the Gaussian, Laplace, and EM can be utilized in both DP and LDP contexts.

 \vspace{-0.1cm}
\begin{table*}[h]
\centering
\caption{Fundamental perturbation mechanisms for differentially private FL} 
 \vspace{-0.1cm}
\begin{tabular}{cll}
\hline
\multicolumn{1}{l}{Data Types} & \multicolumn{1}{c}{DP}                                                                                   & \multicolumn{1}{c}{LDP}                                                                                                          \\ \hline
Continuous                    & Gaussian \cite{dwork2014algorithmic}                                                                     & Laplace \cite{dwork2006calibrating}, Duchi \cite{duchi2013local}, Harmony \cite{nguyen2016collecting}, PM \cite{wang2019collecting} \\
Discrete                      & Discrete Gaussian \cite{wang2020d2p}, Skellam \cite{agarwal2021skellam} & GRR \cite{wang2017locally}, RAPPOR \cite{erlingsson2014rappor}, EM \cite{mcsherry2007mechanism} \\ \hline
\end{tabular}
\label{table:Fundamental perturbation mechanisms for DP-FL}
\end{table*}
\vspace{-0.2cm}

\vspace{-0.2cm}
\subsection{Perturbation Mechanisms for DP in FL} 
In DP within FL, the Gaussian mechanism is most commonly used. Although the Laplace mechanism was originally defined within the context of DP, it is not widely utilized in this scenario. Subsequently, some discrete variants of the Gaussian mechanism have been widely applied in DP-FL, such as Discrete Gaussian and Skellam.

\textbf{Gaussian Mechanism~\cite{dwork2014algorithmic}} is the most popular mechanism for DP in FL. Assume $f:x \to \mathbb{X}$ be a function related to a query. For any $x \in \mathbb{X}$, this mechanism adds noise $n \sim \mathcal{N}(0,\sigma^2)$ to the $f(x)$ to ensure $(\epsilon,\delta)$-DP. Here $\mathcal{N}(0,\sigma^2)$ denotes the noise sample from the Gaussian (normal) distribution with mean $0$ and variance $\sigma^2$, where $\sigma^2=\frac{2\Delta f^2\log(1.25/\delta)}{\epsilon^2}$ and $\Delta f$ is the sensitivity of $f$.

More and more mechanisms are beginning to perform secure aggregation to satisfy differential privacy through encryption. The requirement for encryption to be performed over finite fields has led to the proposal of discrete noise mechanisms. 

{\bf Discrete Gaussian Mechanism \cite{wang2020d2p}} has been widely adopted for this purpose, operating by adding noise drawn from a Discrete Gaussian distribution. Assume $f:x \to \mathbb{X}$ be a function related to a query with sensitivity is 1. For any $x \in \mathbb{X}$, the Discrete Gaussian mechanism adds noise $N_ \mathbb{L} (0,\sigma^2)$ to the $f(x)$ to ensure $(\alpha, \alpha/(2 \sigma ^2))$-RDP. Here $N_ \mathbb{L} (0,\sigma^2) $ is Discrete Gaussian distribution with mean 0 and variance $\sigma^2$ as follows.

\vspace{-0.2cm}
\begin{definition}({\bf Discrete Gaussian Distribution \cite{canonne2020discrete}}).
Let $\mu \in \mathbb{Z}, \sigma \in \mathbb{R}$ with $\sigma \geq 0$. The discrete Gaussian distribution with mean $\mu$ and variance $\sigma^2$ is denoted $N_{\mathbb{L}}(\mu, \sigma ^2)$. It is a probability distribution supported on the integers and defined by
\begin{align}
    \forall x\in\mathbb{Z},\quad\mathbb{P}_{X\sim \mathcal{N}_\mathbb{L}(\mu,\sigma^2)}[X=x]=\frac{e^{-(x-\mu)^2/2\sigma^2}}{\sum_{y\in\mathbb{Z}}e^{-(y-\mu)^2/2\sigma^2}}.
\end{align}
\end{definition}
\vspace{-0.2cm}

{\bf Skellam Mechanism \cite{agarwal2021skellam}} is a response to the issue of the Discrete Gaussian mechanism not being closed under summation. Let $\epsilon > 0, \alpha>1$ and $f:x \to \mathbb{X}$ be a function related to a query. For any $x \in \mathbb{X}$, this mechanism adds noise $n \sim \mathrm{Sk}_{0,\mu}$ to $f(x)$ to satisfy $(\alpha, \frac{\alpha\Delta f^2}{2\mu}+\min\left(\frac{(2\alpha-1)\Delta f^2+6\Delta f}{4\mu^2},\frac{3\Delta f}{2\mu}\right))$-RDP. Here $\Delta f$ is the sensitivity of $f$  and $\mathrm{Sk}_{0,\mu}$ denotes sampling from the Skellam distribution with mean $0$ and variance $\mu$ as follows.

\vspace{-0.2cm}
\begin{definition}({\bf Skellam Distribution \cite{skellam1946frequency}}).
The multidimensional Skellam distribution $Sk_ {\Delta, \mu}$ over $\mathbb{Z} ^ d$ with mean $\Delta \in \mathbb{Z} ^ d$ and variance $\mu$ is given with each coordinate $X_i$ distributed independently as
\begin{align}
    X_i\sim\operatorname{Sk}_{\Delta_i,\mu}\textit{with} \; P(X_i=k)=e^{-\mu}I_{k-\Delta_i}(\mu),
\end{align}
where $ I_v (x)$ is the modified Bessel function of the first kind.
\end{definition}
\vspace{-0.2cm}

In addition to the above, there are Binomial Mechanism~\cite{agarwal2021skellam} and Poisson Binomial Mechanism~\cite{chen2022poisson} also are discrete noise mechanisms for DP in FL, which we will not delve into here due to space constraints.

 \vspace{-0.2cm}
\subsection{Perturbation Mechanisms for LDP in FL}
In LDP within FL, there are three fundamental mechanisms: Laplace Mechanism, Randomized Response (RR), and Exponential Mechanism (EM). The first two are typically used for value perturbation, with the former introducing unbounded noise and the latter bounded noise. The last one is commonly used for value selection, such as dimension selection in FL scenarios. We will first introduce these three basic mechanisms and then discuss some popular and advanced mechanisms built upon them.

\textbf{Laplace Mechanism}~\cite{dwork2006calibrating} originates from data publication scenarios in DP, typically used for mean estimation. Given private data $x \in \mathbb{X}$, this mechanism adds noise $n \sim \text{Lap}(0,\frac{\Delta f}{\epsilon})$ to the aggregated value $f(x)$ of the private data, ensuring $\epsilon$-LDP. Here, $\Delta f$, referring to the $L_1$ sensitivity of $f$, is defined as $\Delta f = \max_{\|x-y\|_1=1} \|f(x)-f(y)\|_1$. In the FL scenario, $f(x)$ typically denotes the local FL model that outputs gradient updates.

\textbf{Randomized Response (RR)}~\cite{warner1965randomized} originates from data collection scenarios in LDP, typically used for count estimation of discrete values. This mechanism typically involves two steps: perturbation and calibration. In the perturbation step, each user perturbs their binary value $x\in\{0,1\}$ probabilistically according to Eq.(\ref{eq:rr_perturb}), ensuring $\epsilon$-LDP. Assuming there are $n$ users' perturbed data, in the aggregation step, the aggregator collects the perturbed count of 1 as $c$, and the actual count of 1 can be derived by calibration via $\hat{c} = \frac{c(e^\epsilon+1) - n}{e^\epsilon - 1}$.
\begin{equation} \label{eq:rr_perturb}
    \Pr[\mathcal{R}_{RR}(x) =v] = \left\{
    \begin{array}{lcr}
    \frac{e^\epsilon}{e^\epsilon + 1}    & \text{if } v = x  \\
    \frac{1}{e^\epsilon + 1}    & \text{if } v = 1-x
    \end{array} \right.
\end{equation}

\textbf{Exponential Mechanism (EM)}~\cite{mcsherry2007mechanism} is used for discrete element selection. Given any utility function (also known as a scoring function), $u$, defined for each value, if the randomized algorithm $\mathcal{M}$ outputs result $r$ with a probability $\mathcal{M}_{E}(x, u, \mathcal{R})\sim \exp \left(\frac{\epsilon u(x, r)}{2 \Delta u}\right)$, then $\mathcal{M}$ satisfies $\epsilon$-LDP.

In the context of mean estimation with continuous data during FL, except for the Laplace mechanism and RR, there are several advanced mechanisms.

\textbf{Duchi's Mechanism}~\cite{duchi2013local} performs numeric mean estimation based on RR. Intuitively, it first randomly rounds the value $x\in[-1,1]$ to a discrete value $v \in \{-1,1\}$, then applies RR and calibrates it. Taken together, this mechanism perturbs $x\in[-1,1]$ to a value $\mathcal{R}_{Duchi}(x) \in \left\{\frac{e^\epsilon + 1}{e^\epsilon - 1}, -\frac{e^\epsilon + 1}{e^\epsilon - 1}\right\}$ according to the probability defined in Eq.(\ref{eq:duchi_perturb}), where the output range of this mechanism is $\{-1,1\}$ times the calibration factor. The aggregator derives the mean by aggregating the sum of users' perturbed values. 
\begin{equation} \label{eq:duchi_perturb}
    \Pr[\mathcal{R}_{Duchi}(x) =v] = \left\{
    \begin{array}{lcr}
    \frac{1}{2} + \frac{e^\epsilon - 1}{2e^\epsilon + 2} \cdot x    & \text{if } v = \frac{e^\epsilon + 1}{e^\epsilon - 1}  \\
    \frac{1}{2} - \frac{e^\epsilon - 1}{2e^\epsilon + 2} \cdot x     & \text{if } v = -\frac{e^\epsilon + 1}{e^\epsilon - 1}
    \end{array} \right.
\end{equation}

\textbf{Harmony}~\cite{nguyen2016collecting} based on Duchi's mechanism, focuses on multi-dimensional vectors. For a bit vector with $d$ dimensions, each user samples only one bit to perturb with Duchi's mechanism, replacing the output range from $\mathcal{R}_{Duchi}(x) \in \left\{\frac{e^\epsilon + 1}{e^\epsilon - 1}, -\frac{e^\epsilon + 1}{e^\epsilon - 1}\right\}$ to $\mathcal{R}_{Duchi}(x) \in \left\{\frac{e^\epsilon + 1}{e^\epsilon - 1}\cdot d, -\frac{e^\epsilon + 1}{e^\epsilon - 1}\cdot d \right\}$.

\textbf{Piecewise Mechanism (PM)~\cite{wang2019collecting}} combines the advantages of Laplace mechanism and Duchi's mechanism by perturbing the private value $x$ to a range $[l(x), r(x)]$ with a larger probability. To ensure $\epsilon$-LDP, the range is defined as $l(x)=\frac{e^{\epsilon/2}\cdot x - 1}{e^{\epsilon / 2} - 1}$ and $r(x)=\frac{e^{\epsilon/2}\cdot x + 1}{e^{\epsilon / 2} - 1}$. The perturbation follows Eq.(\ref{eq:pm_perturb}).
\begin{equation} \label{eq:pm_perturb}
    \Pr[\mathcal{R}_{PM}(x) =v] = \left\{
    \begin{array}{lcr}
    \frac{e^{\epsilon/2}}{2}\cdot \frac{e^{\epsilon/2}-1}{e^{\epsilon/2} + 1}   & \text{if } v \in [l(x), r(x)]  \\
    \frac{1}{2e^{\epsilon/2}}\cdot \frac{e^{\epsilon/2}-1}{e^{\epsilon/2} + 1}    & \text{if } v \notin [l(x), r(x)]
    \end{array} \right.
\end{equation}

In the context of count estimation with discrete values, we list the advanced mechanisms as follows.

\textbf{Generalized Randomized Response (GRR)} \cite{wang2017locally} is an extension of traditional RR from binary values to discrete values in a domain of size $d$. Each user perturbs the value according to Eq.(\ref{eq:grr_perturb}). The aggregator then calibrates the perturbed count $c_x$ of value $x$ as $\hat{c}_x = \frac{c_x(e^\epsilon+d-1) - n}{e^\epsilon - 1}$.
\begin{equation} \label{eq:grr_perturb}
    \Pr[\mathcal{R}_{RR}(x) =v] = \left\{
    \begin{array}{lcr}
    \frac{e^\epsilon}{e^\epsilon + d - 1}    & \text{if } v = x  \\
    \frac{1}{e^\epsilon + d - 1}    & \text{if } v = 1-x
    \end{array} \right.
\end{equation}

\textbf{RAPPOR}~\cite{erlingsson2014rappor}, also building on RR, incorporates a Bloom filter to facilitate discrete frequency estimation in large domains. Specifically, given a discrete domain $\{1,2,\ldots,d\}$, denoted as $[d]$, and $k$ hash functions $\mathbb{H}=\{H_1, H_2, \dots, H_k\}$ that map values from $[d]$ to $[k]$. Each user first uses these $k$ hash functions to map their value $v$ into the Bloom filter with $m$ bits and then perturbs each bit in the Bloom filter using RR. To reduce collisions, where two values are hashed to the same set of indices, RAPPOR allocates users into several cohorts, assigning a unique group of hash functions to each cohort. For longitudinal privacy, which protects against attacks during multiple accesses, RAPPOR splits $\epsilon$ into two parts: the first for permanent perturbation via RR and the second for instantaneous perturbation based on the permanent perturbation.

\end{document}